\documentclass[11pt]{article}
\usepackage[margin=1in]{geometry}
\usepackage[T1]{fontenc}
\usepackage[utf8]{inputenc}
\usepackage{amsmath,amssymb,amsfonts}
\usepackage{graphicx}
\usepackage{xcolor}
\usepackage[numbers,sort&compress]{natbib}
\usepackage{placeins}
\usepackage[colorlinks=true,allcolors=blue]{hyperref}
\usepackage{titlesec}
\titlelabel{\thetitle.\quad}

\begin{document}

\begin{flushleft}
{\huge\bfseries Selective sparsity-enhanced synchronization in disordered semiconductor laser networks\par}
\vspace{1.2em}
{\scshape\large Li-Li Ye,\textsuperscript{1} Nathan Vigne,\textsuperscript{2} Fan-Yi Lin,\textsuperscript{2,3} Hui Cao,\textsuperscript{2,*} and Ying-Cheng Lai\textsuperscript{1,4}\par}
\vspace{0.8em}
{\small\itshape
\textsuperscript{1}School of Electrical, Computer and Energy Engineering, Arizona State University, Tempe, Arizona 85287, USA\\
\textsuperscript{2}Department of Applied Physics, Yale University, New Haven, Connecticut 06520, USA\\
\textsuperscript{3}Institute of Photonics Technologies, Department of Electrical Engineering, National Tsing Hua University, Hsinchu 30013, Taiwan\\
\textsuperscript{4}Department of Physics, Arizona State University, Tempe, Arizona 85287, USA\\[2pt]
\textsuperscript{*}To whom correspondence should be addressed. E-mail: hui.cao@yale.edu\par}
\end{flushleft}

\vspace{0.5em}
\noindent\textbf{Abstract:} Biological networks tend to leverage selective sparsity as an economic strategy to optimize function while minimizing wiring costs. In contrast, achieving synchronization in engineered systems such as semiconductor laser arrays is traditionally expected to require resource-intensive coupling 
to overcome intrinsic frequency disorders. We demonstrate that selectively coupled sparse networks can outperform fully connected architectures, achieving near-complete synchronization with a significantly reduced coupling budget. Using an evolutionary algorithm, we identify a ``pairing opposites" principle: optimal structures specifically prioritize connections between oscillators with the largest opposite frequency detuning. This topology neutralizes dynamical interference induced by redundant links, leading to stable synchronization. We formalize this mechanism through a thermodynamic potential framework, mapping time-delayed phase dynamics to an energy landscape where optimal sparsity prunes additional states to stabilize global phase-locking. Furthermore, we show that the optimal connectivity scales inversely with system size. This principle provides a resource-efficient blueprint for synchronizing and stabilizing diverse complex networks, from photonic arrays to neuromorphic hardware.

\section{Introduction}

Synchronization, the spontaneous alignment of phases and frequencies among interacting oscillators, is a fundamental phenomenon underlying the collective behavior of diverse systems, ranging from neural circuits in the brain to the stable operation of national power grids~\cite{acebron2005kuramoto,arenas:2008,motter2013spontaneous,bullmore2012economy}. In these complex networks, organization is often shaped by an economical trade-off: minimizing ``wiring costs" while maximizing functional efficiency~\cite{bullmore2012economy,barthelemy2011spatial,tero2010rules,mihara:2022,suarez2021learning}. For example, recent studies in neuroscience~\cite{bullmore2012economy,suarez2021learning,naud2018sparse,achterberg2023spatially} suggest that brain networks leverage selective sparsity as a dual-optimization strategy, reducing metabolic costs while simultaneously enhancing information transmission by neutralizing dynamical interference. However, in engineered systems such as semiconductor laser arrays, the prevailing paradigm for overcoming intrinsic frequency disorder has long prioritized increasing the coupling strength~\cite{ye2026disorder,barioni2025interpretable,NHBWB:2021,song2024phase,nixon2011synchronized,arenas:2008,kozyreff2000global}. It is intuitively expected that coupling most lasers, such as in an all-to-all topology, would yield the best global frequency- and phase-locking; yet, its implementation becomes prohibitively complex and energy-intensive as arrays scale up~\cite{yang2023review,glova2003phase,fan2005laser,acebron2005kuramoto,nixon2013observing}. Consequently, simpler topologies such as nearest-neighbor coupling are experimentally common~\cite{orenstein1992large,glova2003phase,nair:2018phase}, but it is difficult to lock frequency outliers in a large array. This raises a fundamental question that transcends specific physical platforms: Can selective sparsity, reflecting the economic organization of biological networks, be engineered to not only save resources but to actually enhance synchronization beyond what is achievable in a fully connected network?

Laser synchronization, both frequency and phase locking of coupled emitters to generate intense coherent beams, serves as a premier testbed for complex and time-delayed network dynamics~\cite{soriano2013complex,sciamanna2015physics,nyaupane2025coherence}. Despite the development of various schemes, ranging from master-slave injection locking~\cite{mercadier2026synchronization,brunner:2015,pfluger:2023} and interaction via a common intermediate~\cite{mahler2020experimental} to utilizing non-Hermitian coupling~\cite{longhi2018invited,liu2022complex,gao2023two}, supersymmetry~\cite{hokmabadi2019supersymmetric,qiao2021higher}, and topological insulator laser arrays~\cite{bandres2018topological,dikopoltsev2021topological}, a major obstacle to large-scale synchronization remains: the inherent frequency detuning among individual lasers. For non-identical Kuramoto oscillators, the synchrony-optimized coupling architectures were identified and the stability was analyzed~\cite{mikaberidze2025emergent,brede2008synchrony,lei2023new}. Coordinated removal of network links was shown to stabilize synchronization by making the completely synchronized state the sole attractor, leading to sparsity-driven synchronization~\cite{mihara:2022}. These findings suggest the possibility of optimizing the connectivity of a sparse laser network for global synchronization. However, applying these general network principles to semiconductor lasers encounters formidable physical complexities absent from traditional oscillator models—most notably, intrinsic amplitude-phase coupling (the linewidth enhancement factor) and time-delayed interactions.
Therefore, the question remains outstanding of whether an optimized sparse network structure exists for global synchronization of semiconductor lasers. 

In this work, we demonstrate that selectively sparse laser networks can paradoxically outperform all-to-all coupling, even when operating with a significantly reduced coupling budget. Employing an island-based genetic algorithm~\cite{cantu1998survey} to evolve the network topology, we find that synchronization can be maximized at an optimal level of sparsity. Our results reveal a counterintuitive principle: optimal structures instead strategically prioritize connections between lasers with the largest frequency difference. Dense connectivity is not merely energetically expensive but dynamically counterproductive, as redundant links between oscillators with mismatched frequencies inject ``frustration'' into the system’s dynamic landscape in phase space. By ``pairing opposites", the network effectively neutralizes the primary sources of frequency disorder by selectively connecting lasers with opposite detunings from the mean frequency. 

To formalize this mechanism, we develop a thermodynamic potential theory by recasting the time-delayed phase dynamics of the Lang-Kobayashi (LK) equations~\cite{lang:1980,alsing:1996} as a gradient flow. This theoretical framework reveals that the steady states of synchronization correspond to local minima in a potential landscape. Notably, the theory predicts that optimal sparsity scales inversely with the number of oscillators, a scaling law that provides a universal and resource-efficient blueprint for stabilizing large-scale disordered systems. Beyond photonics, this principle of sparsity-enhanced synchronization offers a robust route for designing complex networks where connectivity is a premium, from neuromorphic hardware~\cite{kudithipudi2025neuromorphic,li2026pruning} to synthetic biological oscillators~\cite{hollo2025intracellularly}.

\begin{figure} [ht!]
\centering
\includegraphics[width=\linewidth]{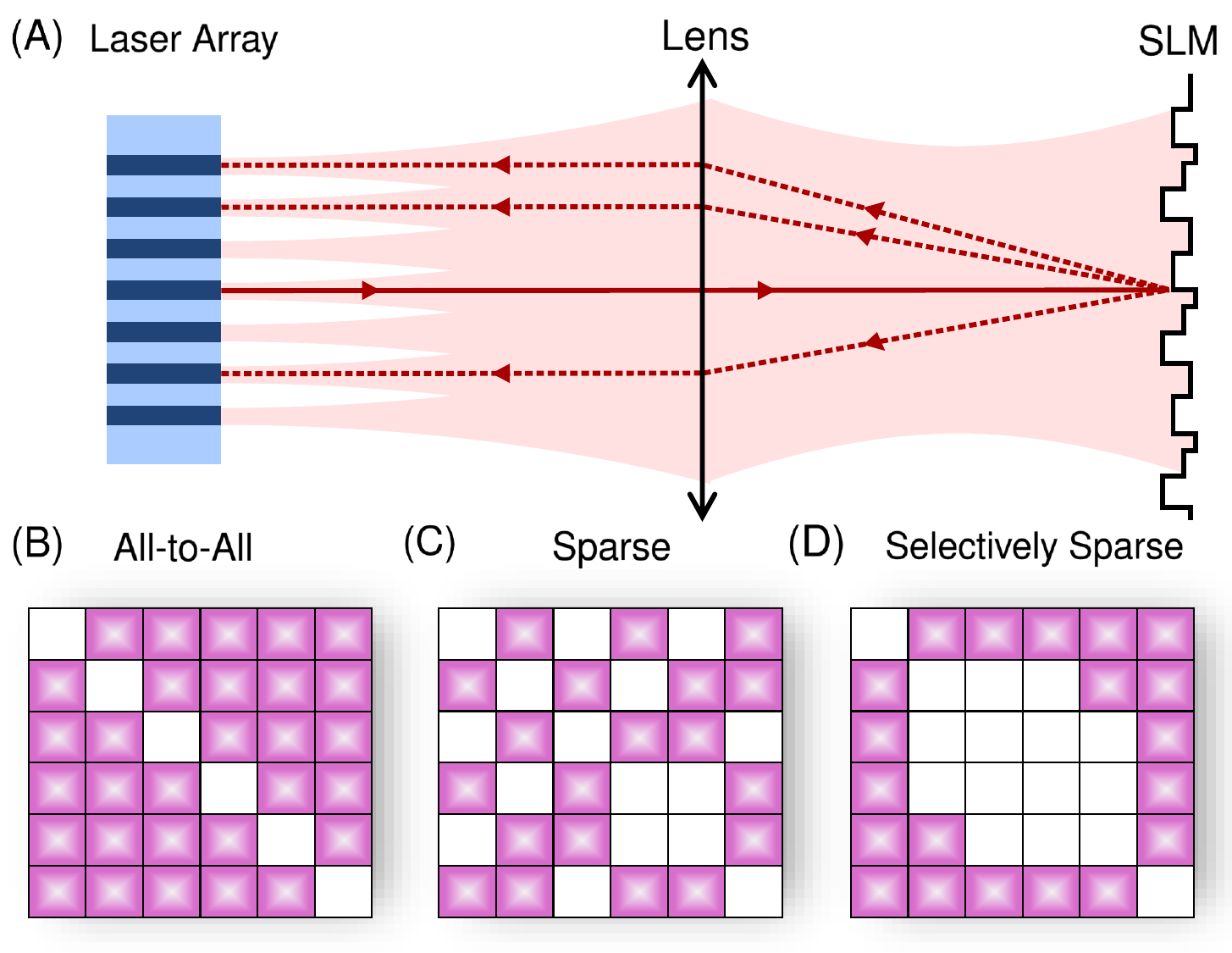}
\caption{Schematic diagram of laser coupling architectures. (A) Programmable laser coupling via re-configuring diffraction patterns of the SLM. (B–D) Laser coupling matrix with different coupling structures for synchronization in the presence of intrinsic frequency disorder: all-to-all, randomly sparse, and selectively sparse configurations, where both the horizontal and vertical axes denote the laser No., and a colored (blank) block at $(p,q)$ indicates that lasers $p$ and $q$ are coupled (uncoupled).}
\label{fig:schematic}
\end{figure}

\section{Results} \label{sec:results}

\subsection{Semiconductor lasers with time-delayed coupling}
	
Figure~\ref{fig:schematic}(A) illustrates a laser array coupled externally through a spatial light modulator (SLM). The array is at the front focal plane of an optical lens and the SLM is placed at the back focal plane. The lasers at different transverse locations are coupled via the gratings of corresponding periods on the SLM. The coupling is delayed by the round-trip time $\tau$ between the SLM and the laser array. Light reflection from the SLM forms an external cavity for the lasers. 
In the absence of the SLM, each laser operates in a single longitudinal and transverse mode. Their intrinsic lasing frequencies deviate randomly due to the inevitable heterogeneity generated, e.g., during the manufacturing process.
	
A network of externally coupled semiconductor lasers with gain saturation and amplitude-phase coupling~\cite{winful1988stability} is mathematically described by the LK equations~\cite{lang:1980}:
\begin{align} \nonumber
	\frac{dE_{p}(t)}{dt}&=\frac{1+i\alpha}{2}\left(g\frac{N_{p}(t)-N_{0}}{1+s|E_{p}(t)|^{2}}-\gamma\right)E_{p}(t)+ i\Delta_p E_{p}(t) \\ \label{eq:LK_1}
&  + e^{-i\omega_{0}\tau}\sum^{L}_{q=1}K_{pq}E_{q}(t-\tau)+F_{E_{p}}, \\ \label{eq:LK_2}
\frac{dN_{p}(t)}{dt}&=J_{0}-\gamma_{n}N_{p}(t)-g\frac{N_{p}(t)-N_{0}}{1+s|E_{p}(t)|^2}|E_{p}(t)|^2 +F_{N_{p}},
\end{align}
where $E_{p}(t)$ is the complex electric field of the $p$th laser and $N_{p}(t)$ is its carrier number. The pump rate $J_0=4J_{\rm th}$ is four times above the lasing threshold, $J_{\rm th}=\gamma_{n}(N_{0}+\gamma/g)$. Here, $\gamma_{n}$, $N_0$, $\gamma$, and $g$ represent the carrier decay rate, the carrier number at transparency threshold, the cavity loss rate, and the differential gain coefficient, respectively. $\alpha$  is the linewidth enhancement factor that characterizes the amplitude-phase coupling, $s$ describes gain saturation, $\tau$ is the time delay of laser coupling. The natural frequency detuning of the $p$th laser is $\Delta_p = \sigma_{\Delta}\,\mathcal{N}(0,1)$, where $\sigma_{\Delta}$ characterizes the magnitude of frequency disorder and $\mathcal{N}(0,1)$ is a Gaussian random variable with zero mean and unit variance. The reference $\omega_0$ of the natural angular frequencies is chosen to satisfy $\omega_0 \tau = 2n\pi$, where $n$ is an integer and selected to be closest to the mean of all laser frequencies. For convenience, we number the lasers by their frequency detuning from low to high: $\Delta_1\leq\Delta_2\leq...\leq\Delta_L$. (This numbering scheme is not related to the spatial distance between the lasers.) Spontaneous emission is introduced into the LK equations as dynamic noise for the electric field, $\langle F_{E_{p}}(t), F^{*}_{E_{q}}(t')\rangle=R_{sp}\delta_{pq}\delta(t-t')$, where $R_{sp}$ is the spontaneous emission rate. Stochastic fluctuations in carrier dynamics are given by $\langle F_{N_{p}}(t), F_{N_{q}}(t')\rangle =\gamma_{n}N_{p}(t)\delta_{pq}\delta(t-t')$.

With external coupling, the electric field of the $p$th laser is $E_p(t)=r_p(t) e^{i\Omega_p(t)}$, where $r_p(t)$ and $\Omega_p(t)$ are the time-varying amplitude and phase, respectively. The short-term frequency $\tilde{\Delta}_p$ is extracted from linear fitting $\Omega_p(t) = \tilde{\Delta}_p \, t + \varphi_p(t)$ over a short time window around $t$, and $\varphi_p(t)$ is phase fluctuation. Complete synchronization encompassing field amplitude, frequency, and phase is quantified by the order parameter $ S=\langle |\sum_{p=1}^L E_p(t)|^2/[L\sum_{p=1}^L|E_p(t)|^2]\rangle_t \in[0,1]$~\cite{NHBWB:2021}, where $S=1$ corresponds to all lasers having identical field amplitude, frequency and phase.  When all amplitudes are equal, $S$ reduces to the standard phase-synchronization parameter which accounts only for phase coherence. Here we focus on $S$, because the uniform amplitude and phase will result in the most constructive interference of all lasers in the far field, producing the strongest intensity peak which is demanded for coherent beam combining. Note that typical phase locking requires only the relative phase between lasers to be constant. A zero relative phase corresponds to in-phase locking. In particular, in-phase-locked lasers will produce a single lobe of highest intensity in the far field via constructive interference, which is often needed for applications. 

\begin{figure*} [ht!]
\centering
\includegraphics[width=0.9\linewidth]{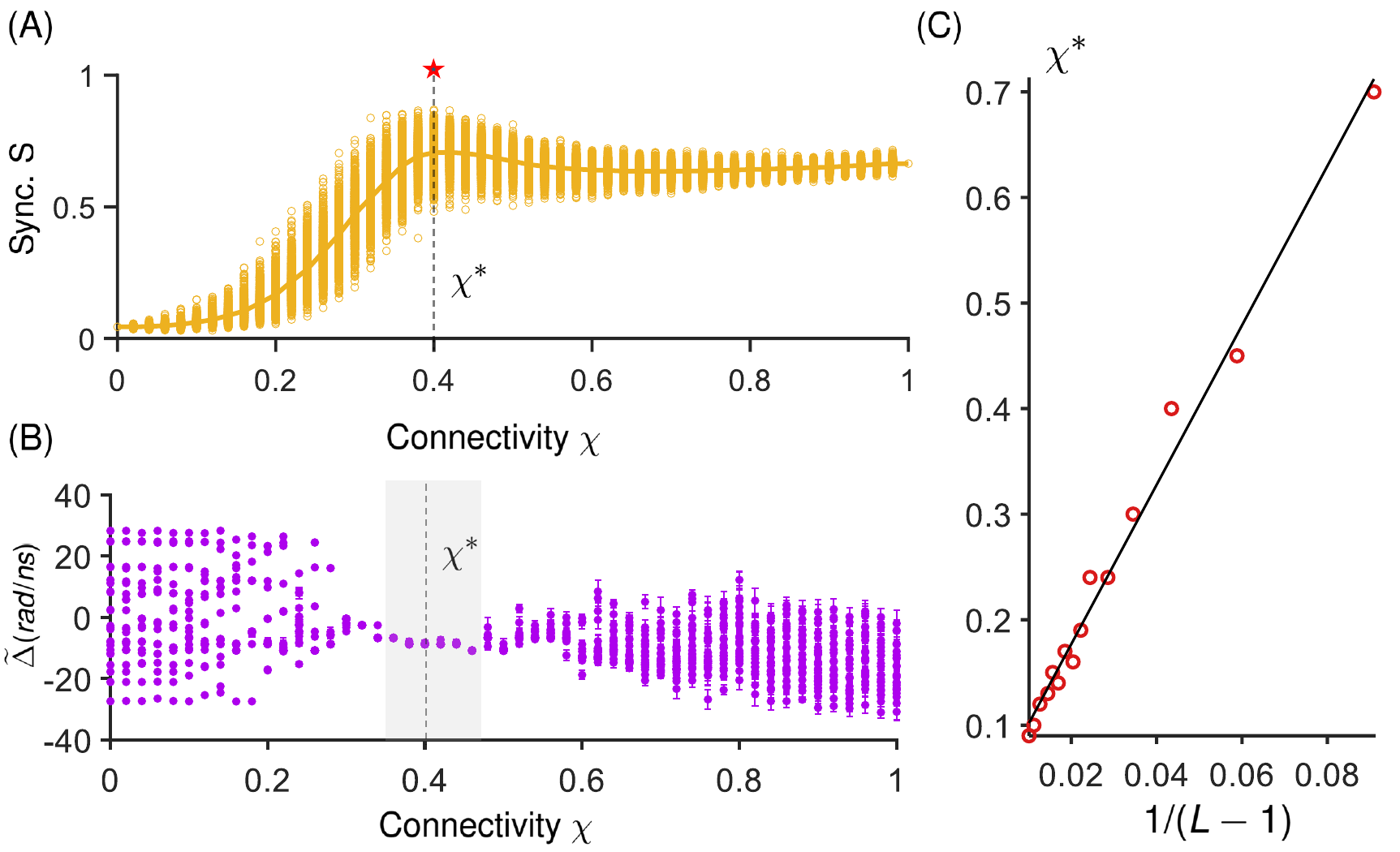}
\caption{Sparsity-enhanced synchronization and scaling of the optimal connectivity. (A) Selective sparsity-driven synchronization of 24 semiconductor lasers with frequency disorders. Orange circles represent the order parameter $S$ for 1000 randomly sparse networks with different connectivity $\chi$, the orange curve is the average for each $\chi$. The red star denotes an order parameter of $S \approx 0.96$, achieved by the optimal selectively sparse network found via the genetic algorithm. (B) Final frequencies of 24 lasers in the randomly sparse network yielding the highest $S$ for each $\chi$ in (A). These frequencies are linearly fitted within moving windows of size $\tau$ with a step size of $0.1\tau$, and the vertical bars indicate the resulting standard deviations. All lasers are frequency locked within the shaded region.  (C) The optimal connectivity $\chi^{*}$ scales as $1/(L-1)$, where $L$ is the number of lasers. Red circles are numerical data, and black line is a linear fit. The values of parameters in the LK equations are $\gamma_{n}  = 0.5\,\textnormal{ns}^{-1}$, $N_{0}=1.5\times 10^{8}$, $\gamma=500\,\textnormal{ns}^{-1}$, $g=1.5\times\,10^{-5}\,\textnormal{ns}^{-1}$,  $\alpha=5$, $s=2\times 10^{-7}$, and $\tau= 3\,\textnormal{ns}$.}
\label{fig:connect_scaling}
\end{figure*}

\begin{figure*} [ht!]
\centering
\includegraphics[width=0.8\linewidth]{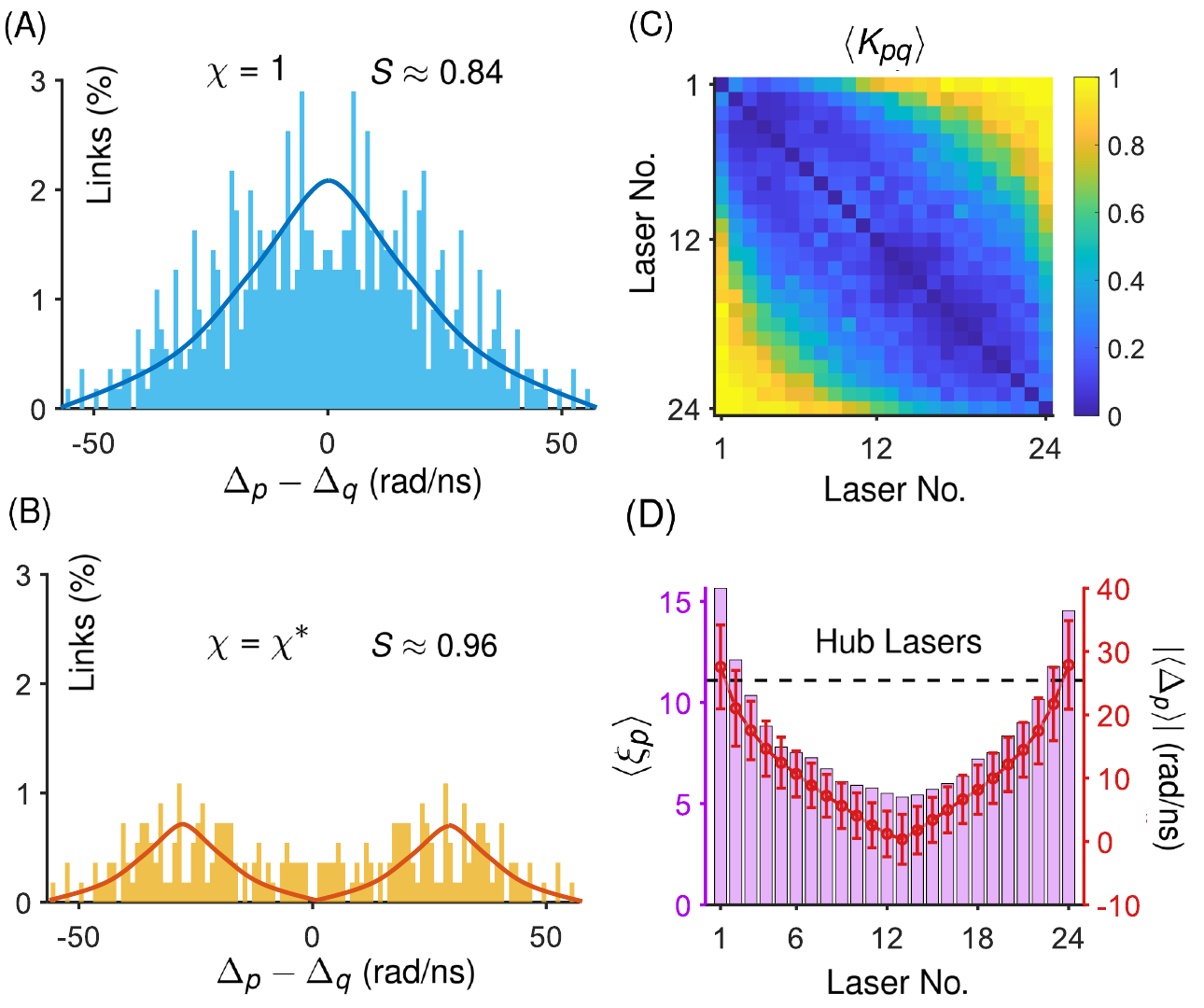}
\caption{Emergence of hub structure in the optimized sparse laser network. (A) Histogram of the percentage of links between lasers as a function of their frequency differences $\Delta_{p}-\Delta_{q}$ ($p,q = 1,\ldots, L$) for homogeneous all-to-all coupled networks. (B) Same legend as in (A) but for the optimized sparse laser networks found by the genetic algorithm at the optimal connectivity $\chi=\chi^{*}$. Solid curves show the histogram are single-peaked in (A) and double peaked in (B), revealing the links between lasers with small frequency difference are mostly removed in optimized sparse networks.  (C) Heat maps of the ensemble average $\langle K_{pq}\rangle$. The average is taken over $100$ optimal selectively sparse coupling topologies $K_{pq}$, each of which is specifically optimized for a distinct realization of frequency disorder. (D) Histogram of the average number of connections per laser, defined as $\langle\xi_{p}\rangle=\sum_q\langle K_{pq}/\kappa^{f}\rangle$ and calculated from the matrix elements in (C), for the optimized sparse networks. The horizontal dashed line indicates the baseline given by the mean value, $\mu_{\langle\xi\rangle}=\sum_p \langle\xi_p\rangle/L$, plus one standard deviation, $\sigma_{\langle\xi\rangle}$. Hub lasers are defined as those with $\langle\xi_{p}\rangle$ exceeding this baseline. Also shown on the right-hand side is the absolute value of the initial frequency detuning from the mean for each laser averaged over $100$ frequency disorder realizations, with standard deviations indicated by the vertical bars.}
\label{fig:hub}
\end{figure*}
\subsection{Sparsity-enhanced synchronization}

To generate randomly sparse networks of lasers, we start from an all-to-all coupled network [Fig.~\ref{fig:schematic}(B)] of lasers with identical coupling coefficient and randomly or selectively remove some links [Fig.~\ref{fig:schematic}(C,D)]. Thus, the coupling coefficients have binary values of $K_{pq}=0$ or $1\, \textnormal{ns}^{-1}$, and self-feedback is neglected ($K_{pp}=0$). The network connectivity is defined as the ratio between the total number of connected links and the maximum possible number of links: 
\begin{align}
\chi \equiv \frac{\sum_{p<q}K_{pq}/\kappa^{f}}{L(L-1)/2}
\end{align}
with $K_{pq}$ normalized by the coupling strength, $\kappa^{f}=1\,{\rm ns^{-1}}$, to count the number of links. We simulate the dynamics of the sparse network for $t\in[0,100]$ ns and calculate the order parameter $S$ for $t\in[50,100]$ ns, well beyond the transient dynamics. 
The orange circles in Fig.~\ref{fig:connect_scaling}(A) represent $S$ values for individual networks, and the orange curve is the ensemble average of $S$, $\langle S \rangle_e$, in different random configurations with the same connectivity $\chi$. As $\chi$ gradually increases from zero to one, $\langle S \rangle_e$ first increases relatively rapidly and then decreases. The peak of $\langle S \rangle_e$ gives the optimal connectivity $\chi^* = 0.4$. It reveals that removing links from a fully connected network ($\chi=1$) can actually improve synchronization with a reduced coupling budget. This result even holds for laser arrays with inhomogeneous coupling strength (see Sec.~\ref{sec:SI:inhomogeneous} in the SI). Furthermore, this behavior is consistently observed across different frequency disorders and initial seeds (see Sec.~\ref{sec:SI:Optimizing} in the SI). 

To find the range of network connectivity for frequency locking, we compute the final frequencies of individual lasers in the time window $[50,100]$ ns. For each value $\chi$ in Fig.~\ref{fig:connect_scaling}(A), we select the coupling configuration that yields the maximum $S$. Figure~\ref{fig:connect_scaling}(B) shows that in the vicinity of $\chi^{*}$, all laser frequencies merge, confirming frequency locking. Below this regime (shaded), $\chi$ is too weak for frequency locking; above this regime, the final lasing frequencies exhibit large fluctuations, indicating that the coupling is so strong that it induces chaotic dynamics and reduces $S$. This dynamical explanation parallels our previous results on disorder-mediated synchronization resonance in homogeneous all-to-all-coupled networks~\cite{ye2026disorder}.

We further find that the optimal connectivity scales with the number of lasers as $\chi^{*}\propto 1/(L-1)$, as shown in Fig.~\ref{fig:connect_scaling}(C). The slight deviation arises from the statistical fluctuations of the frequency disorder. A theoretical explanation of this scaling law is provided below.


\subsection{Optimized sparse network}

Figure \ref{fig:connect_scaling}(A) shows large variations in the order parameters for different coupling configurations of identical connectivity near $\chi^*$. This suggests that there may exist some sparse networks that achieve near-complete synchronization. To find such optimized sparse networks, we select the network connectivity at $\chi^*$ with $K_{pq} = 0~\textnormal{or}~1~\rm{ns^{-1}}$, and use an island-based genetic algorithm to search for the optimal coupling configuration that gives the highest possible value of $S$. 
The red star in Fig. \ref{fig:connect_scaling}(A) marks $S\approx$ 0.96 for
an optimized sparse network of $L$ = 24 lasers with $\chi^*$ = 0.4 found by a genetic algorithm. The network optimization process is repeated for 100 statistically independent realizations with frequency disorder. All reach near-perfect synchronization and the ensemble-averaged order parameter is $\langle S\rangle_e\approx 0.97 \pm 0.01$. Furthermore, the optimized results for L=100 lasers are detailed in Sec.~\ref{sec:SI:Optimizing} of the SI.

\begin{figure*} [ht!]
\centering
\includegraphics[width=
\linewidth]{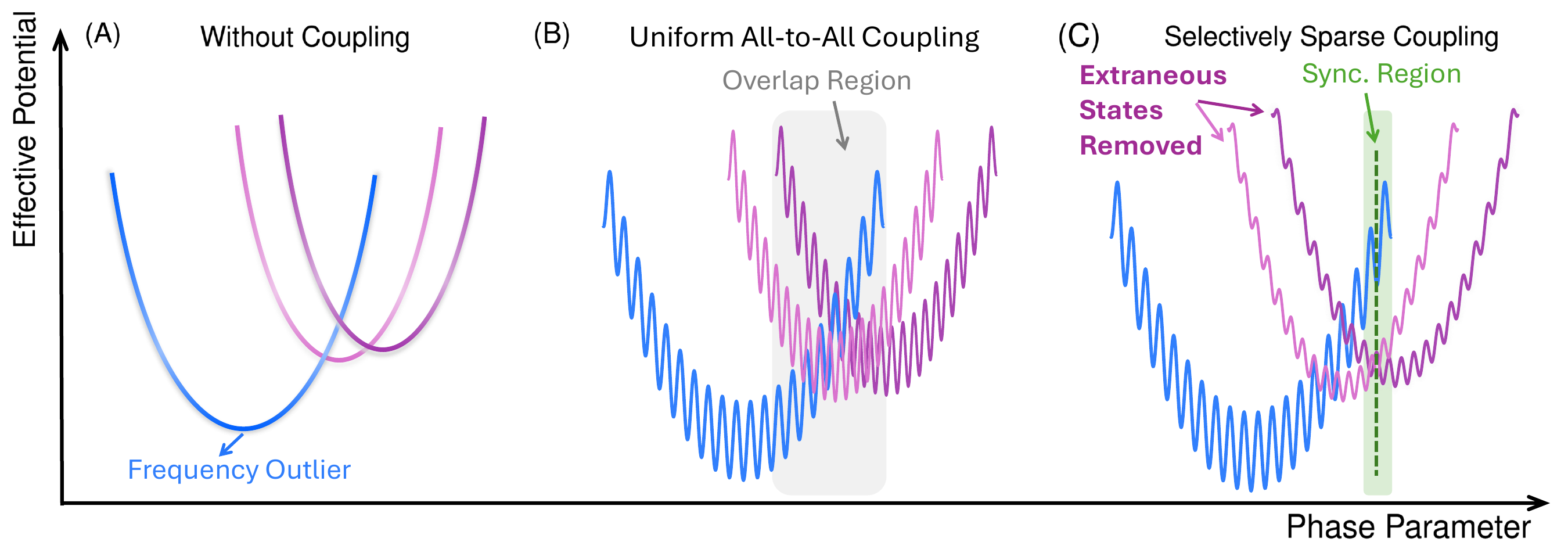}
\caption{Effect of selectively sparse coupling on the thermodynamic effective potential. (A) In the absence of coupling, the potential landscape over the phase parameter space (the time-delayed phase difference of the electric field) consists of independent, smooth quadratic potentials. Frequency disorders shift these global minima apart, as illustrated by three curves with a distinct frequency outlier (blue curve) separated from the rest of the array (light and dark purple curves). (B, C) Uniform all-to-all (B) and selectively sparse (C) coupling reshape the landscape via sinusoidal modulations, creating multiple local minima. Gray and green shaded areas highlight the overlap and synchronized regions, where modulated potentials exhibit significant energy barriers at the same phase, pulling disparate steady-state solutions together. Notably, sparse coupling (C) eliminates extraneous steady states at small frequency disorder (purple curves) because off-center energy barriers are too low. While narrowing the synchronized overlap region, it enhances dynamic stability relative to all-to-all coupling (B). The 
dashed line in (C) indicates the typical steady-state synchronization.}
\label{fig:theory}
\end{figure*}

To understand how the sparse networks is optimized for synchronization, we examine which links are removed from all-to-all coupled networks. Figures~\ref{fig:hub}(A,B) show the percentage of coupled links between lasers with the natural frequency difference $\Delta_{p}-\Delta_{q}$ in fully connected and optimized sparse networks. The former has more links for smaller $|\Delta_{p}-\Delta_{q}|$, because natural frequencies have a Gaussian distribution and are more likely to be close to each other. In the latter, most links between lasers with small frequency differences are removed. Figure~\ref{fig:hub}(C) shows the ensemble average of the optimized coupling matrix across 100 frequency disorders, confirming that lasers with larger frequency differences are more likely to be coupled. This can be further seen from Fig.~\ref{fig:hub}(D), the number of links for each laser, where the lasers with the largest frequency offset from the mean have the highest number of links, so they are effectively ``hub'' lasers. Note that they are different from the hub in a star network~\cite{fischer2006zero,zamora2010crowd,bourmpos2012sensitivity,cohen2012phase}, which mediates the coupling among lasers and serves as the ``bridge'' for synchrony. In our optimized sparse network, the hub lasers have the largest frequency detuning and are most difficult to be locked in frequency. As a result, more connections of such lasers to other lasers, especially those with opposite frequency detuning, are needed to pull the hub laser frequencies to the mean.   


The synchronization behavior remains consistent with and without noise in our numerical simulation (see Sec.~\ref{sec:SI:Robust:InitialNoise} in the SI). The selectively-coupled networks identified by the island-based genetic algorithm also exhibit robustness to perturbations in laser frequency detuning, coupling strength, initial states, and pump rate, highlighting the diversity and resilience of the genetic-algorithm landscape (Secs.~\ref{sec:SI:Robust:InitialNoise}, \ref{subsec:SI:Robust:FreqCoupPump} and \ref{sec:SI:Robust:GA} in the SI). Furthermore, we numerically show that synchronization remains stable under sudden and gradual turn-on of laser coupling (Sec.~\ref{subsec:SI:Robust:ramping} in the SI).

\subsection{Theoretical modeling} \label{sec:understanding}

Intuitive insights into the benefits of selective sparsity and hub-based topologies can be gained by a simplified linear oscillator model (Sec.~\ref{sec:SI:LinearNetwork} in the SI). With increasing connectivity, the lowest-loss coupled state (supermode) expands much more quickly in a network with selective coupling than random coupling. Once the supermode spreads over the entire network, global synchronization is possible.


To acquire a more quantitative understanding, we have developed a theoretical framework to identify the relationship between sparse network topology and the solution space of synchronized steady-states under frequency disorders, as detailed in Secs.~\ref{sec:SI:Thermo} and \ref{sec:SI:theory} in the SI. In particular, we apply thermodynamic potential theory to determine the steady-state solutions for each laser, utilizing an effective potential that incorporates interactions from other lasers. Generally, each coupled laser possesses a distinct set of steady-state solutions; as the coupling strength increases, the number of these solutions typically grows. A completely synchronized steady-state occurs when these individual solutions  for all lasers overlap entirely on the potential landscape. While frequency disorders push the steady-state solutions of
individual lasers apart, specific classes of network topologies, most notably hub-emergent selectively sparse networks, can pull them together, as shown in Fig.~\ref{fig:theory}.

In the absence of coupling (Fig.~\ref{fig:theory}(A)), the LK equations for an individual laser can be reduced to phase dynamics~\cite{alsing:1996} characterized by a gradient flow. By invoking the steady-state condition and the slow-phase-evolution approximation~\cite{lenstra:1991}, this flow is shown to be governed by a thermodynamic potential possessing a well-defined global minimum, typically a smooth quadratic potential. Under these dynamics, any arbitrary initial state undergoes a transient collapse, eventually stabilizing at this minimum. Consequently, the global minimum uniquely defines the steady-state solution for the isolated laser. Frequency disorders shift these global minima within the solution space, pushing the individual steady-state solutions apart and preventing synchronization across the array.

The coupled network reshapes the potential landscape by introducing appreciable sinusoidal modulations within each laser's solution space~\cite{lenstra:1991}, as shown in Figs.~\ref{fig:theory}(B,C). This coupling introduces multiple local minima partitioned by energy barriers~\cite{NHBWB:2021,ye2026disorder}. These emergent local minima with appreciable energy barriers represent alternative steady-state solutions for each individual laser. Notably, as the distance of these new solutions to the global minimum decreases, the associated energy barriers heighten, thereby enhancing the local stability of these states. Moreover, the height of the energy barriers near the global minimum can be modulated by the total coupling strength per laser $k^{\textnormal{in}}_p=\sum_qK_{pq}$. For each laser, as the total coupling strength per laser $k^{\textnormal{in}}_p$ becomes stronger, more stable steady-state solutions emerge, which may eventually destabilize the dynamics and drive the system into a chaotic regime~\cite{winful:1990}. 

Increased the total coupling strength per laser $k^{\textnormal{in}}_p$ expands the set of stable steady-state solutions for each laser, facilitating phase-matching with the collective steady state. However, this creates a trade-off: while a higher density of solutions promotes synchronization, it also risks driving the dynamics into a chaotic regime. For individual interacting pairs, the amplitude-phase coupling ($\alpha$ factor) and external time delay introduce an intrinsic phase shift between coupled lasers, mathematically analogous to the frustration parameter in the Sakaguchi-Kuramoto model~\cite{sakaguchi1986soluble, yeung:1999}. In a densely connected network with heterogeneous frequencies, this inherent phase frustration scales into collective dynamical frustration. Here, dynamical frustration~\cite{daido1992quasientrainment,barre2016bifurcations} emerges when competing interactions from the dense coupling topology and over-constrained phase-matching requirements driven by the $\alpha$ factor and time delay $\tau$ prevent the system from relaxing into a single global steady state. Instead, these conflicting constraints fracture the potential landscape into a rugged terrain characterized by multistability and chaos. Here, multistability refers to the coexistence of multiple stable steady-state solutions for the same set of parameters. Within this fractured landscape, the emergent local minima act as auxiliary steady states to facilitate phase-matching. 

While lasers with minimal frequency detuning require few such auxiliary states, frequency outliers depend on these extended solutions at the expense of dynamical stability. In this regard, by pruning extraneous solutions for smaller frequency detunings (Fig.~\ref{fig:theory}(C)), a sparse network effectively neutralizes this dynamical frustration and stabilizes the system more effectively than uniform all-to-all coupling, as shown in Fig.~\ref{fig:connect_scaling}(A), Figs.~\ref{fig:hub}(B-D) and Figs.~\ref{fig:theory}(B,C). This implies that the total coupling strength per laser, $k^{\textnormal{in}}_p$ is proportional to the frequency disorder $\Delta_p$, as $k^{\textnormal{in}}_p\propto\Delta_p$. Such a relationship results in the emergent hub structure shown in Fig.~\ref{fig:hub}(D), where the degree of each laser, defined as $\xi_p = k^{\textnormal{in}}_p / \kappa^f$, is proportional to the frequency disorder $\Delta_p$.

Moreover, as detailed in Secs.~\ref{sec:SI:Thermo} and \ref{sec:SI:theory} of the SI, the physical scaling of the network can be understood through the reduction of the individual total coupling strengths per laser, $k^{\text{in}}_p$, to a global critical coupling threshold, $k_c^{\text{in}}$. Physically, global synchronization emerges the moment the network's collective coupling ``pull'' becomes strong enough to capture the most extreme frequency outlier. This transition occurs when the effective global coupling perfectly balances the maximum frequency detuning. Since the requisite pulling force (the global critical coupling $k_c^{\text{in}}$ for each laser) is determined by this largest detuning ($k_c^{\text{in}} \propto \max|\Delta_p|$), the total coupling cost scales linearly with the number of lasers $L$ ($\sum_p k_c^{\text{in}} \propto L \max|\Delta_p|$). Here, $\max|\Delta_p|$ is approximately constant because it grows very slowly with the network size $L$, scaling as $\sqrt{\ln L}$~\cite{gumbel2004statistics} due to the tail properties of the Gaussian distribution. Consequently, the resulting critical connectivity $\chi_c$ scales inversely with system size $L$, $\chi_c \propto 1/(L-1)$, as further detailed in Sec.~\ref{sec:SI:theory} of the SI. Increasing the connectivity further generates multiple near-synchronized steady-state solutions, corresponding to the periodic orbit phase, or the edge of chaos. The optimal connectivity $\chi^*$ lies within this phase, where $\chi^* \gtrsim \chi_c\propto1/(L-1)$, as shown in Fig.~\ref{fig:connect_scaling}(C). 

\section{Discussion and conclusion}

In the optimized sparse network, lasers with minimal frequency detuning do not require direct connections, as their indirect coupling through hub lasers is sufficient for synchronization. As the number of lasers $L$ increases, the optimal networks become progressively sparser because these indirect coupling paths multiply. Consequently, optimal connectivity scales to $\chi^{*} \propto 1/(L - 1)$, indicating that the coupling budget increases linearly with the system size as $\sum_{pq}K_{pq}\propto L$. This linear scaling, fundamentally replacing the prohibitively expensive quadratic scaling with $L$ of dense networks, facilitates large-scale synchronization. These results provide a principled, universal approach to designing scalable, cost-effective networks that sustain robust frequency and phase synchronization, along with a stable steady state, even under significant frequency disorder.  In this work, only real coupling coefficients with constant amplitude are considered, both the amplitude and phase of the coupling coefficients can be further optimized, which will be pursued in future studies (see Sec.~\ref{sec:SI:Thermo} in the SI).

In conclusion, we demonstrated that selectively sparse networks can paradoxically outperform dense, all-to-all architectures in synchronizing disordered laser arrays. By combining numerical optimization through an island-based genetic algorithm with a thermodynamic potential framework, we revealed that dense connectivity is not merely resource-intensive but dynamically counterproductive. Instead, we uncovered a ``pairing opposites'' principle: selectively linking oscillators with the largest opposite frequency detuning effectively neutralizes dynamical frustration. This finding highlights that selective sparsity, a known organizational principle in biological connections like the ``economy of brain networks,"~\cite{bullmore2012economy} is not merely a biological necessity but a profound and robust physical strategy for complex systems. By pruning redundant links that induce chaotic instabilities, our approach bridges the gap between biological evolution and the rational design of large-scale, cost-effective photonic hardware.

Specifically, the island-based genetic algorithm constructs these optimized sparse networks by adaptively varying the number of connected links per laser to compensate for the frequency disorder. Beyond mere disorder compensation, this sparsity reduces the overall coupling cost $\sum_{pq}K_{pq}$, thereby suppressing multistability and promoting robust frequency- and phase-locked dynamics. We elucidated the physical mechanism for the emergence of such sparse networks and their hub structures. While the critical coupling strength in the classical Kuramoto model with homogeneous all-to-all coupling has been well characterized~\cite{chopra:2009,jadbabaie:2004, dorfler:2011}, we generalized this analysis by incorporating the linewidth enhancement factor, time delay, and sparse coupling. Furthermore, we reformulated the optimal connectivity from a physical perspective using a thermodynamic potential framework. Beyond semiconductor lasers, our results can be extended to solid-state lasers and fiber lasers as well as other classes of coupled nonlinear oscillators. 


\section*{Data, Materials, and Software Availability}
The simulated data are available at Zenodo: \url{https://zenodo.org/records/21653378}. The code for reproducing the results presented in this work is available on GitHub: \url{https://github.com/liliyequantum/Optimal-Sparse-Network-Lasers}.


\section*{Acknowledgement}
This work was supported by the Office of Naval Research under Grant No.~N00014-24-1-2548.

\section*{Author contributions and competing interests}
H.C. and Y.-C.L. designed research; L.-L.Y. performed research; L.-L.Y., N.V., F.-Y.L., H.C., and Y.-C.L. analyzed data; and L.-L.Y., N.V., H.C., and Y.-C.L. wrote the paper. The authors declare no competing interest.

\bibliographystyle{unsrt}
\bibliography{refs}

\clearpage

\setcounter{section}{0}
\setcounter{subsection}{0}
\setcounter{figure}{0}
\setcounter{equation}{0}
\renewcommand{\thesection}{S\arabic{section}}
\renewcommand{\thesubsection}{\thesection\Alph{subsection}}
\renewcommand{\thefigure}{S\arabic{figure}}
\renewcommand{\theequation}{S\arabic{equation}}

\begin{center}
{\LARGE\bfseries Supplementary Information}\\[10pt]
{\large Selective sparsity-enhanced synchronization in disordered semiconductor laser networks}\\[6pt]
Li-Li Ye, Nathan Vigne, Fan-Yi Lin, Hui Cao, and Ying-Cheng Lai
\end{center}

This Supplementary Information (SI) provides supporting material for the main text. In Sec.~\ref{sec:SI:inhomogeneous}, we present the phenomenon of synchronization resonance in inhomogeneous networks and compare it with that arising in homogeneous all-to-all coupled networks, demonstrating that the resonance is robust to the network structure. In Sec.~\ref{sec:SI:Optimizing}, we examine the phenomenon of connectivity resonance in sparse networks in more detail and explain how genetic algorithm was used to find the optimal network structure (Sec.~\ref{subsec:SI:Optimizing:GA}). We demonstrate that the optimized networks show reduced preference for connections between lasers with close frequency detuning. Furthermore, the temporal dynamics of the synchronized steady state in the $L=100$ laser array are presented. In Sec.~\ref{sec:SI:Robust}, we investigate the robustness of synchronization in sparse networks against dynamic noise, initial-state perturbations, and random variations in frequency detuning, coupling strength, pump rate, and coupling ramping. Furthermore, we mix multiple optimized sparse coupling matrices to explore the optimization landscape of the genetic algorithm to address the issue of landscape robustness. In Sec.~\ref{sec:SI:LinearNetwork}, we simplify the Lang-Kobayashi equations to obtain a network of coupled linear oscillators to gain intuition for the optimal sparse network structure, providing a quick guide for experimental interpretation. In Sec.~\ref{sec:SI:Thermo}, we provide a more quantitative understanding by analytically deriving the effective thermodynamic potential from the full Lang-Kobayashi equations for the system of coupled-lasers with time delay caused by the external-cavity feedback. In Sec.~\ref{sec:SI:theory}, we provide a detailed physical understanding of the optimal sparse network structure, including the selective sparsity-enhanced synchronization, the emergence of hub structures and the scaling law of optimal connectivity. 

\section{Synchronization resonance in inhomogeneous all-to-all coupled laser networks} \label{sec:SI:inhomogeneous}

\begin{figure} [ht!]
\centering
\includegraphics[width=0.8\linewidth]{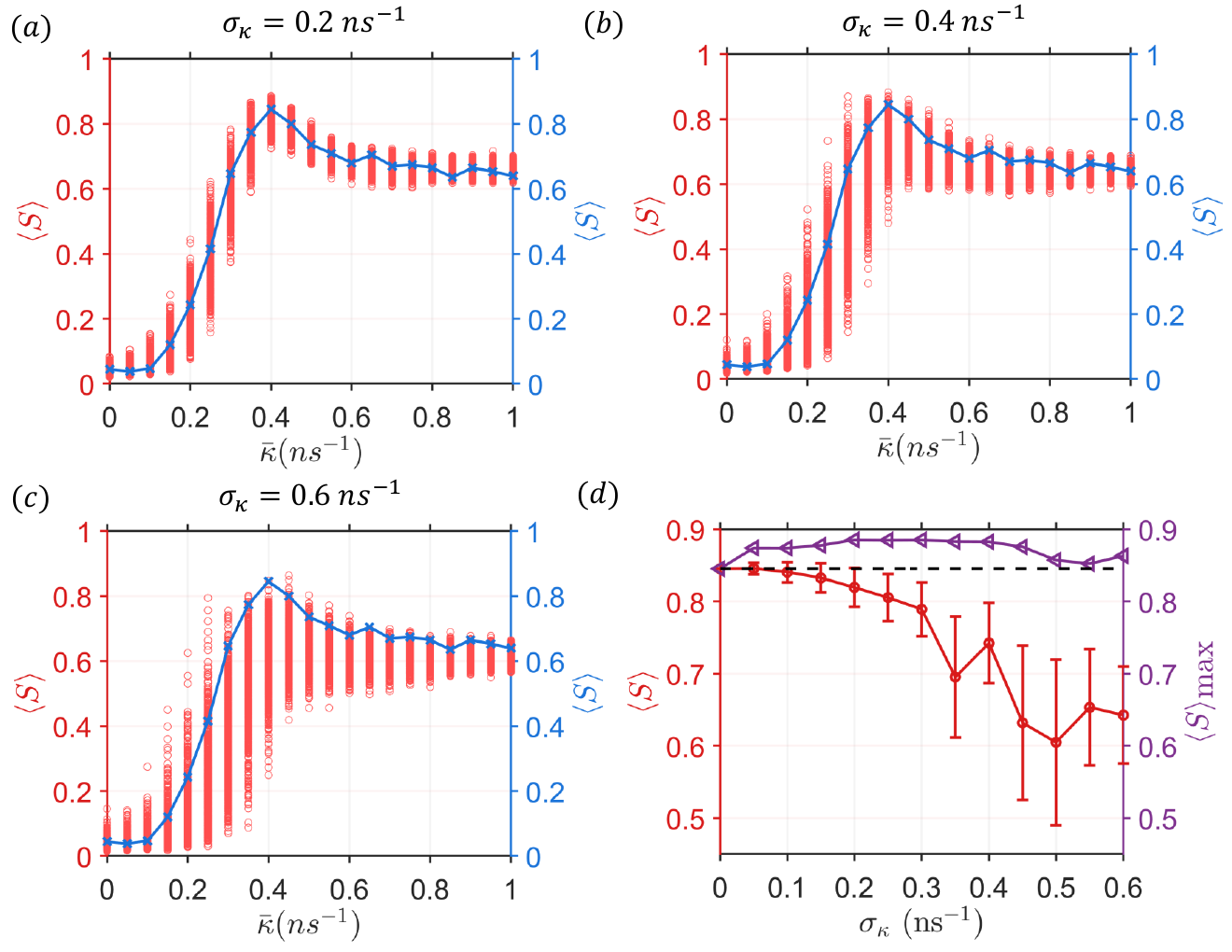}
\caption{Synchronization resonance in inhomogeneous all-to-all coupled diode laser networks. The nodal coupling is given by $K_{pq}=\left[\bar{\kappa}+\sigma_{\kappa}\mathcal{N}(0,1)\right](1-\delta_{pq})$ with $\mathcal{N}(0,1)$ being a normalized Gaussian random variable. (a-c) Synchronization measure $\langle S\rangle$ (For simplicity, the order parameter denoted strictly as $\langle S \rangle$ throughout this Supplementary Information is referred to simply as S in the main text.) versus $\bar{\kappa}$ for $\sigma_{\kappa} = 0.2$ ns$^{-1}$, 0.4 ns$^{-1}$, and 0.6 ns$^{-1}$, respectively, for $L = 24$, where the blue curve represents the baseline from the homogeneous all-to-all coupled network with $K_{pq}=\bar{\kappa}(1-\delta_{pq})$. The coupling strength is varied over the range $\bar{\kappa} \in [0, 1]\,\textnormal{ns}^{-1}$ with a step size of $0.05\,\textnormal{ns}^{-1}$. (d) Mean values and standard deviations of the randomly sampled synchronization measure $\langle S\rangle$ at the resonance peak $\bar{\kappa}^{*}=0.4\,{\rm ns^{-1}}$ versus the perturbation strength $\sigma_{\kappa}$. The corresponding peak values of $\langle S\rangle_{\rm max}$ as a function of $\sigma_{\kappa}$ are displayed on the right. For reference, the horizontal dashed line marks the maximum $\langle S\rangle$ obtained in homogeneous all-to-all networks, providing a baseline for comparison.}
\label{fig:inhomogeneous}
\end{figure}

In homogeneous, all-to-all coupled diode laser networks, a synchronization resonance emerges. The coupling matrix is defined as $K_{pq}=\kappa$ for $p\neq q$ and $K_{pp}=0$, thereby eliminating self-feedback. Under frequency disorder, the synchronization measure $\langle S\rangle$ (For simplicity, the order parameter denoted strictly as $\langle S \rangle$ throughout this Supplementary Information is referred to simply as S in the main text.) reaches a maximum at an optimal coupling strength, whereas in the absence of disorder this resonance disappears. A natural question arises: does synchronization resonance, as exemplified by the blue curves in Figs.~\ref{fig:inhomogeneous}(a–c), persist when the couplings become inhomogeneous? We find that deviations from homogeneous coupling preserve and may even slightly enhance the synchronization resonance.

To introduce coupling heterogeneity, we perturb the interaction strength as 
\begin{align} 
\kappa \rightarrow \bar{\kappa} + \sigma_{\kappa} \mathcal{N}(0,1),
\end{align}
where $\bar{\kappa}$ denotes the baseline coupling and $\mathcal{N}(0,1)$ represents a normalized Gaussian random variable. For each perturbation level $\sigma_{\kappa}$, we generate 1000 independent realizations of the perturbed coupling matrices for every fixed value of $\bar{\kappa}$. Figures~\ref{fig:inhomogeneous}(a–c) present results for $L=24$ lasers with frequency disorders $\Delta_p = 14 \times \mathcal{N}(0,1)$ rad/ns generated using the random seed \texttt{rng(1)}, a typical disorder configuration. The plots show the dependence of $\langle S\rangle$ on $\bar{\kappa}$ for $\sigma_{\kappa} = 0.2$ ns$^{-1}$, 0.4 ns$^{-1}$, and 0.6 ns$^{-1}$, respectively, with the blue curve serving as a reference for the homogeneous network. Despite the spread in $\langle S\rangle$ at each fixed $\bar{\kappa}$, synchronization resonance persists in all cases. Figure~\ref{fig:inhomogeneous}(d) summarizes the mean values and standard deviations of $\langle S\rangle$ at the resonance peak (left) and the corresponding peak values $\langle S\rangle_{\rm max}$ as a function of $\sigma_{\kappa}$ (right). The horizontal dashed line marks the maximum $\langle S\rangle_{\rm max}$ obtained for the homogeneous, all-to-all coupled network. These results demonstrate that random variations in coupling strength can partially compensate for frequency disorder, thereby enhancing overall synchronization.

Based on the theoretical framework described in Sec.~\ref{sec:SI:theory}, inhomogeneous all-to-all coupling introduces small variations in the total coupling strength experienced by each laser. These variations can partially compensate for frequency disorder, thereby reducing the number of possible steady-state solutions. Consequently, the final laser frequencies become more closely aligned, resulting in more stable dynamics and a moderate enhancement of both frequency and phase synchronization.

\begin{figure} [ht!]
\centering
\includegraphics[width=\linewidth]{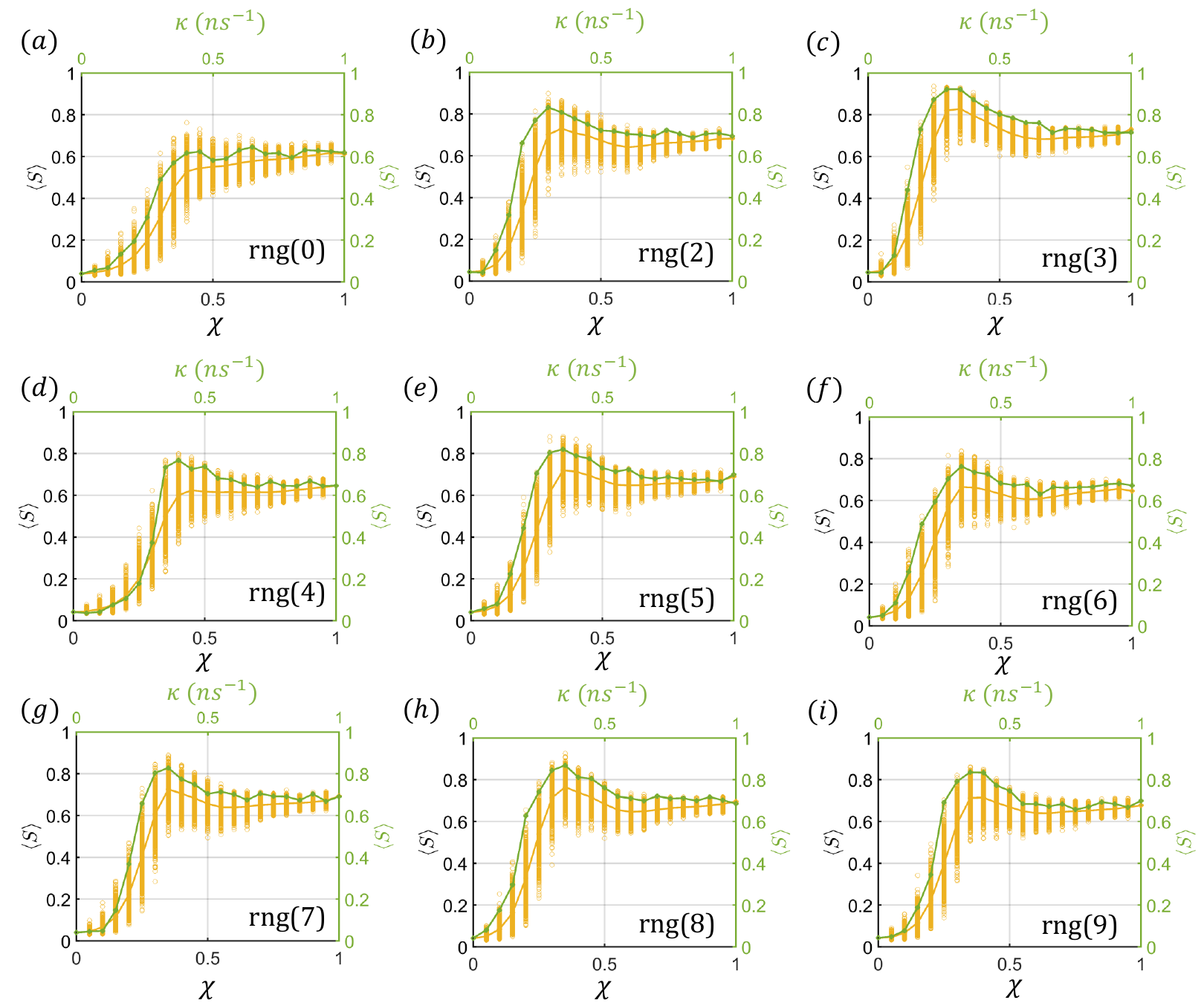}
\caption{Connectivity resonance in sparsely all-to-all coupled diode laser networks. Frequency disorder is modeled as Gaussian detuning, $\Delta_p = 14 \times \mathcal{N}(0,1)\,\textnormal{rad/ns}$, with nine independent realizations (panels a–i, random seeds \texttt{rng(0)}, \texttt{rng(2)}–\texttt{rng(9)}; \texttt{rng(1)} shown in Fig.~\ref{fig:connect_scaling}(A) of the main text). In each panel, the green curve denotes the homogeneous all-to-all benchmark $K_{pq}=\kappa(1-\delta_{pq})$, while 1000 yellow points per connectivity level represent randomly sampled sparse configurations, with the yellow curve indicating their average.}
\label{fig:supp_connect_scan}
\end{figure}

\section{Optimizing synchronization in sparsely coupled diode-laser networks} \label{sec:SI:Optimizing}

\subsection{Connectivity resonance in synchronization} \label{subsec:SI:Optimizing:Connectivity}

The all-to-all coupled diode-laser networks studied in Sec.~\ref{sec:SI:inhomogeneous} demonstrate that random inhomogeneous coupling can partially compensate for frequency disorder and thereby enhance synchronization. A different form of inhomogeneity arises from the network structure itself, specifically, through sparse coupling architectures. From an experimental standpoint, sparsely coupled laser networks offer practical advantages in implementation. As discussed in the main text, network sparsity is quantified by the connectivity $0 \le \chi \le 1$, defined as the fraction of nonzero entries in the coupling matrix normalized by the total number of possible links $L(L-1)$. The case $\chi = 1$ corresponds to all-to-all coupling. As $\chi$ increases from zero, synchronization attains a maximum at an optimal connectivity $\chi^*$, giving rise to the phenomenon of connectivity resonance, as described in the main text. Here we provide additional results illustrating the emergence of this resonance, as shown in Fig.~\ref{fig:supp_connect_scan}.

In our simulations, following experimental considerations, the coupling matrix is constrained to be binary (entries are either 0 or 1), symmetric, and free of self-coupling (zero diagonal elements). For each connectivity level, we perform random sampling over 1000 sparse coupling configurations with different frequency disorder realizations. The resulting values of $\langle S\rangle$ are shown as yellow points in Figs.~\ref{fig:supp_connect_scan}(a–i), corresponding to nine statistically independent realizations of the frequency disorder. In each panel, the green curve represents the homogeneous, all-to-all reference case.

For the homogeneous all-to-all network with coupling strength $\kappa$, the total intrinsic coupling per laser is
\begin{align}
	k_{\textnormal{Homo}}^{\textnormal{in}} = \sum_{q}K_{pq}=\kappa (L-1).
\end{align}
In the case of sparse binary coupling, the expected total coupling per laser is
\begin{align} 
k_{\textnormal{Spa}}^{\textnormal{in}}=\mathbb{E}[\sum_{q}K_{pq}]=\chi (L-1)\kappa^{f}
\end{align}
where $\kappa^{f}=1\textnormal{ns}^{-1}$. For a fixed total coupling per laser, there exists a one-to-one correspondence between $\kappa$ and the connectivity $\chi$. Remarkably, even with randomly sampled configurations, sparse networks can outperform the homogeneous all-to-all networks near the synchronization resonance region, as shown in Figs.~\ref{fig:supp_connect_scan}(a–i). Furthermore, the synchronization peak $\kappa^{*}$ observed in the homogeneous case statistically coincides with the optimal connectivity $\chi^{*}$ in sparse networks, indicating that the synchronization threshold is primarily determined by the total coupling per laser.

\subsection{Exploiting island genetic algorithm to find optimal sparsely coupled networks} \label{subsec:SI:Optimizing:GA}

\begin{figure} [ht!]
\centering
\includegraphics[width=0.9\linewidth]{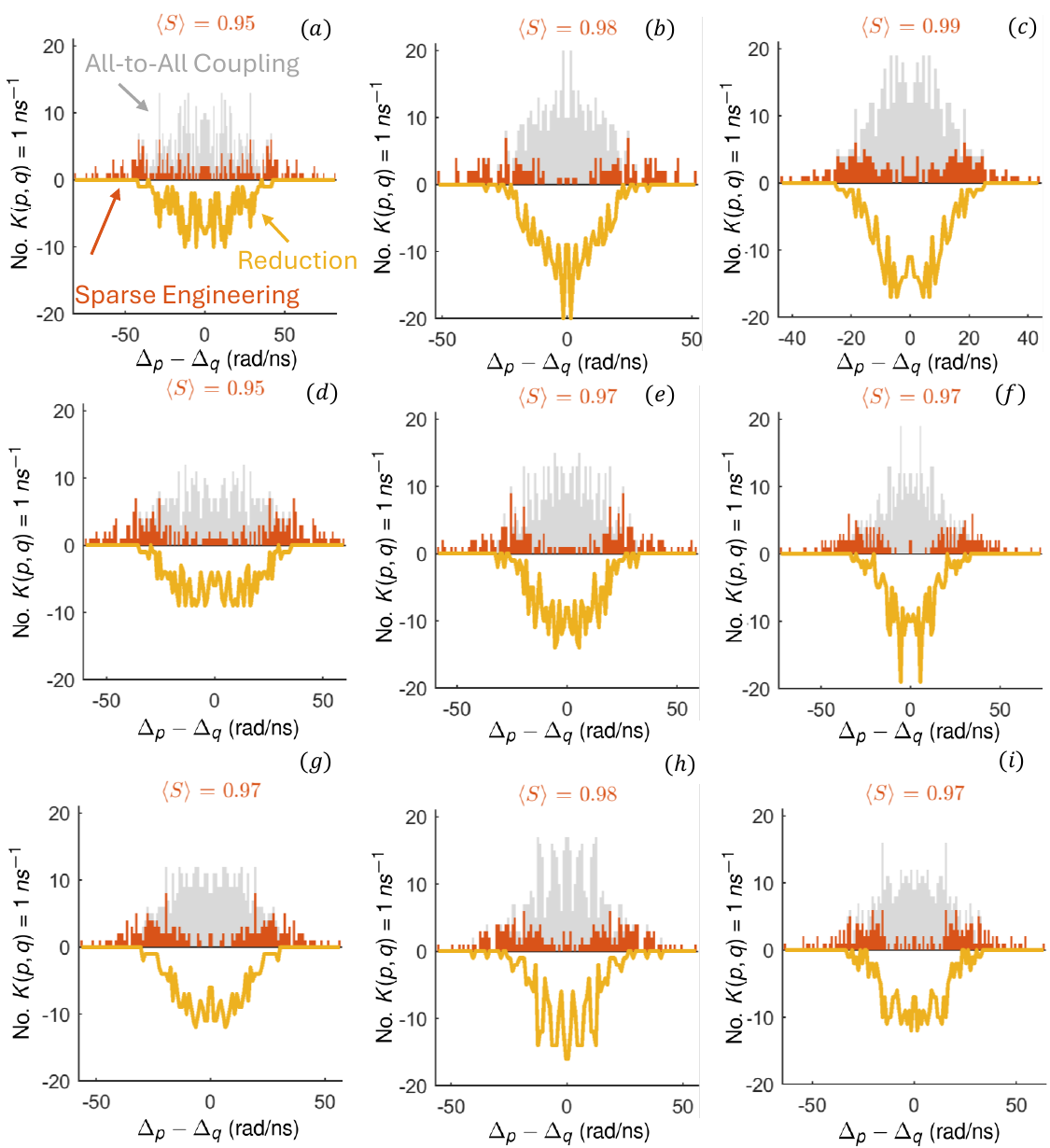}
	\caption{Comparison of coupling characteristics in homogeneous all-to-all networks and in optimal sparse networks identified by an island genetic algorithm. Shown are the statistical distributions of the network couplings as a function of frequency detuning differences. Panels (a–i) correspond to random seeds \texttt{rng}(0) through \texttt{rng}(9), excluding \texttt{rng}(1) (already presented in the main text). The ordering of random seeds matches that in Fig.~\ref{fig:supp_connect_scan}. To emphasize the effect of sparsity, the coupling strength is fixed at $\kappa = 1\,\text{ns}^{-1}$ for the homogeneous all-to-all case, while links are selectively removed using an island genetic algorithm to enhance synchronization. The removed connections indicate nonessential links. In each panel, the gray histogram shows the synchronization distribution for the homogeneous case $\langle S \rangle \approx 0.67$, averaged over all frequency disorder realizations), the orange histogram corresponds to the optimal sparse network, and the yellow histogram highlights the difference between the two.}
\label{fig:supp_freq_dist_kij}
\end{figure}

In Sec.~\ref{subsec:SI:Optimizing:Connectivity}, we showed that Monte Carlo sampling of sparsely coupled configurations can yield networks exhibiting stronger frequency and phase synchronization than any homogeneous all-to-all network with the same total coupling cost $K_{\rm total}$ in the weak-coupling regime. This observation implies the existence of optimal sparse networks with enhanced synchronization, motivating the use of optimization methods such as genetic algorithms to identify such structures. To this end, we employ an island genetic algorithm, initialized at the connectivity level that produces the best synchronization performance in the Monte Carlo simulations, to discover coupling configurations with further improved synchronization.

The implementation of the island genetic algorithm for our network optimization problem proceeds as follows. A standard genetic algorithm is first applied to evolve a population of 200 candidate coupling matrices per generation. The optimization typically converges after approximately 200 generations. Each matrix is evaluated using the synchronization measure $\langle S \rangle$ as the fitness function. Parent matrices are selected via roulette-wheel selection, assigning higher selection probabilities to individuals with higher fitness. Crossover is then performed between selected parent pairs, with each pair generating two offspring. Edges shared between the parents are directly inherited, while the remaining edges are randomly assigned to preserve the total number of links and ensure the same structural constraints as the parents. Each offspring undergoes mutation at a fixed rate of 0.03, in which one randomly chosen link is removed and replaced by a new random connection elsewhere in the network. The top 8 high-performing individuals (elites) are preserved in each generation.

To promote diversity and avoid premature convergence to local optima, we implement a migration strategy across four independent islands, each containing 50 individuals. Every five generations, migration occurs sequentially: Island 1 $\rightarrow$ 2, 2 $\rightarrow$ 3, 3 $\rightarrow$ 4, and 4 $\rightarrow$ 1, where the worst-performing 5\% of individuals on the receiving island are replaced by the elites migrating from the source island. This island-based genetic algorithm framework enhances exploration of the optimization landscape while preserving the best-performing solutions, leading to robust identification of sparse coupling networks with optimal synchronization properties.

The patterns providing physical insight into the optimized sparse coupling matrices obtained from the island genetic algorithm are shown in Figs.~\ref{fig:supp_freq_dist_kij}(a–i) across different frequency disorder realizations corresponding to random seeds \texttt{rng}(0–9). These results reveal a characteristic structure in which couplings are suppressed between lasers with small frequency differences and preserved between those with large frequency separations. Consequently, the resulting histogram exhibits a dip near zero frequency difference and an overall parabolic shape. To isolate the effect of sparsity, coupling matrix elements for both the homogeneous all-to-all and optimized sparse configurations are constrained to binary values, either 0 or $1\,\text{ns}^{-1}$. Figures~\ref{fig:supp_freq_dist_kij}(a–i) present histograms quantifying the number of coupling elements with $K_{pq} = 1\,\text{ns}^{-1}$ within discrete intervals of the frequency differences $\Delta_p - \Delta_q$. In homogeneous all-to-all networks, all off-diagonal elements satisfy $K_{pq} = 1\,\text{ns}^{-1}$, so the histogram of $K_{pq}$ versus $\Delta_p - \Delta_q$ directly reflects the statistical distribution of the frequency differences. As the frequencies follow a Gaussian distribution, the corresponding histogram (gray background in Fig.~\ref{fig:supp_freq_dist_kij}) also appears Gaussian-shaped, with a mean synchronization level $\langle S\rangle\approx 0.67$ averaged over all frequency disorder realizations. In contrast, the optimized sparse networks exhibit a marked reduction in couplings between lasers with small frequency detuning differences  $\Delta_p - \Delta_q \rightarrow 0\,\text{rad/ns}$ , while preferentially preserving connections among those with larger frequency separations. This trend is captured by the yellow curves in Fig.~\ref{fig:supp_freq_dist_kij}, which display a distinct dip near $\Delta_p - \Delta_q \rightarrow 0\,\textnormal{rad/ns}$ and an overall parabolic profile across the full range of frequency differences. These results suggest that optimal sparse networks actively break the symmetric coupling pattern of the homogeneous configuration to exploit frequency heterogeneity: by linking more detuned lasers, the system balances coupling-induced locking and disorder-driven desynchronization, thereby creating an effective mechanism of structural adaptation that promotes stronger global coherence.

\begin{figure*} [ht!]
\centering
\includegraphics[width=\linewidth]{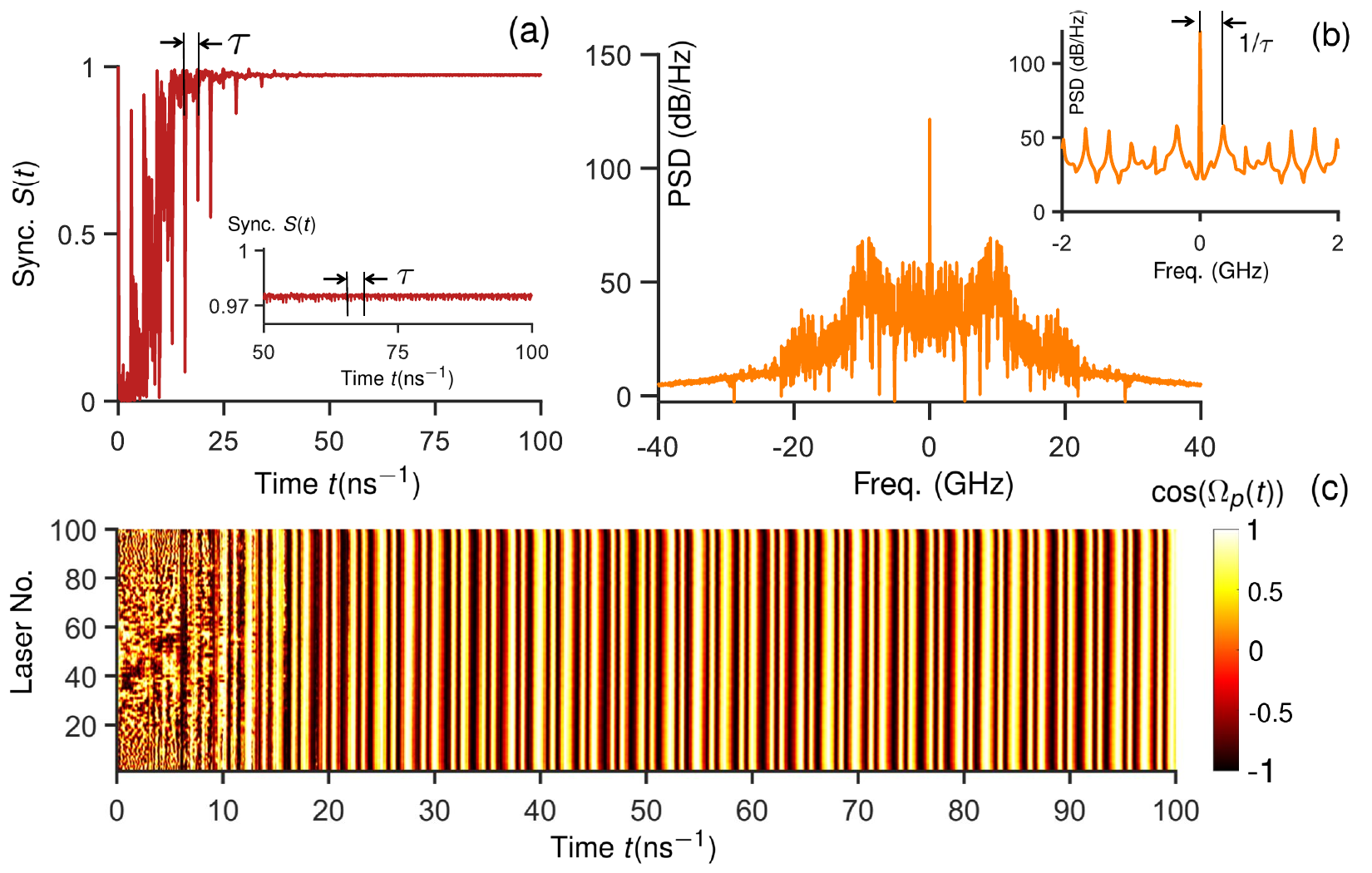}
\caption{The steady-state synchronization in an $L=100$ laser array via selectively sparse coupling. (a) Time series of the order parameter $S(t)$ for the selectively sparse network optimized by the island-based genetic algorithm. (b) Power spectral density (PSD) of the normalized total field intensity during the target time regime. (c) Spatiotemporal evolution of the phase function $\cos\Omega_p(t)$ of the laser field. Parameter values are: pump rate $J_0 = 4\gamma_{n}(N_{0}+\gamma/g) \approx 3.67 \times 10^8 \text{ ns}^{-1}$ and $\sigma_{\Delta} = 14 \text{ rad/ns}$.}
\label{fig:GA_time_series}
\end{figure*}

Figure \ref{fig:GA_time_series} illustrates the optimized steady-state synchronization for an array of $L=100$ lasers via the island-based genetic algorithm. Initial conditions were set to the free-running steady state. (Robustness against perturbations in frequency detuning, the coupling matrix, and initial states, as well as variations in pump rate, time-dependent coupling ramping, and dynamic noise, is detailed in Sec.~\ref{sec:SI:Robust}.). The time evolution of the order parameter $S(t)$ indicates the transition from transient dynamics to a steady state with a fixed fluctuation interval $\tau$, as shown in Fig.~\ref{fig:GA_time_series}(a). The synchronized lasers reach a continuous-wave steady state, where the total intensity $I_{\text{tot}}(t) = |\sum_p E_p(t)|^2$ exhibits negligible temporal fluctuations. This is quantified by the normalized total intensity, $\tilde{I}_{\text{tot}}(t) \equiv (I_{\text{tot}}(t) - \langle I_{\text{tot}} \rangle) / \langle I_{\text{tot}} \rangle$. The corresponding peak at zero frequency in the power spectral density (PSD) of the normalized field intensity $\tilde{I}_{\text{tot}}(t)$ confirms a steady state with constant amplitude. The secondary peaks at approximately $\pm10$ GHz correspond to relaxation oscillations between the carrier and photon densities, which are enhanced by the external cavity feedback. The inset in Fig.~\ref{fig:GA_time_series}(b) reveals the peak interval $1/\tau$, which indicates that the time delay arising from the external cavity affects the oscillation patterns in both the transient and steady-state regimes of the order parameter $\langle S\rangle$ and the total field intensity $\tilde{I}_{\text{tot}}(t)$. Finally, Fig.~\ref{fig:GA_time_series}(c) displays the spatiotemporal evolution of the phase component, $\cos[\Omega_p(t)]$, of the electric field for each laser, confirming frequency and in-phase locking across the array.

\section{Robustness of synchronization in sparse laser networks} \label{sec:SI:Robust}

\subsection{Robustness against dynamic and initial-state noises} \label{sec:SI:Robust:InitialNoise}

Incorporating the dynamic and initial-condition noises into the Lang-Kobayashi equations leads to the following stochastic differential equations~\cite{NHBWB:2021}:
\begin{align} \label{eq:LK_noise_1}
    d E_{p}(t)/dt &=\frac{1+i\alpha}{2}\left(g\frac{N_{p}(t)-N_{0}}{1+s|E_{p}(t)|^{2}}-\gamma\right)E_{p}(t)
    +i\Delta_p E_{p}(t) + e^{-i\omega_{0}\tau}\sum^{L}_{q=1}K_{pq}E_{q}(t-\tau)+F_{E_{p}}, \\ \label{eq:LK_noise_2}
    dN_{p}(t)/dt &=J_{0}-\gamma_{n}N_{p}(t)-g\frac{N_{p}(t)-N_{0}}{1+s|E_{p}(t)|^2}|E_{p}(t)|^2+F_{N_{p}}.
\end{align}
Depending on the wavelength, the natural frequency of the diode lasers lies on the order of $10^2$ THz, which poses challenges for numerical simulation. It is conventional to adopt a rotating frame at the reference frequency $\omega_{0} \equiv 2n\pi/\tau$ for some integer $n$ so that the shift factor $\exp[-i\omega_{0}\tau]$ accounts for the phase accumulation induced by the time delay. Thus, the detuning of the $p$-th laser with true angular frequency $\omega_{p}$ is $\Delta_{p}=\omega_{p}-\omega_{0}$. The steady state of the uncoupled free-running lasers, $E^{s}_{p}(t)=r_{s}\exp[i\Delta_{p} t]$, where $r_{s}$ is the steady-state electric field amplitude and $N_{p}(t)=N_s$, over the interval $t\in[-\tau,0]$ can be set as the initial condition for fast convergence, where 
\begin{align} 
	r^{2}_{s} &=[J_{0}-\gamma_{n}(\gamma/g+N_{0})]/[\gamma(1+\gamma_{n}s/g)], \\ 
	N_{s} &=(1+sr^{2}_{s})\gamma/g+N_{0}.
\end{align}
A major source of dynamic noise in the diode lasers is spontaneous emission in the electric field given by~\cite{NHBWB:2021}
\begin{align} \label{eq:spontaneous_emission}
\langle F_{E_{p}}(t), F^{*}_{E_{q}}(t')\rangle=R_{sp}\delta_{pq}\delta(t-t'),
\end{align}
with $R_{sp}$ being the spontaneous emission noise strength. This is essentially a Gaussian white noise. Another source of dynamic noise is the stochastic fluctuations in the carrier dynamics, which is multiplicative with intensity proportional to the carrier number:
\begin{align} \label{eq:carrier_noise}
\langle F_{N_{p}}(t), F_{N_{q}}(t')\rangle =\gamma_{n}N_{p}(t)\delta_{pq}\delta(t-t').
\end{align}

\begin{figure} [ht!]
\centering
\includegraphics[width=\linewidth]{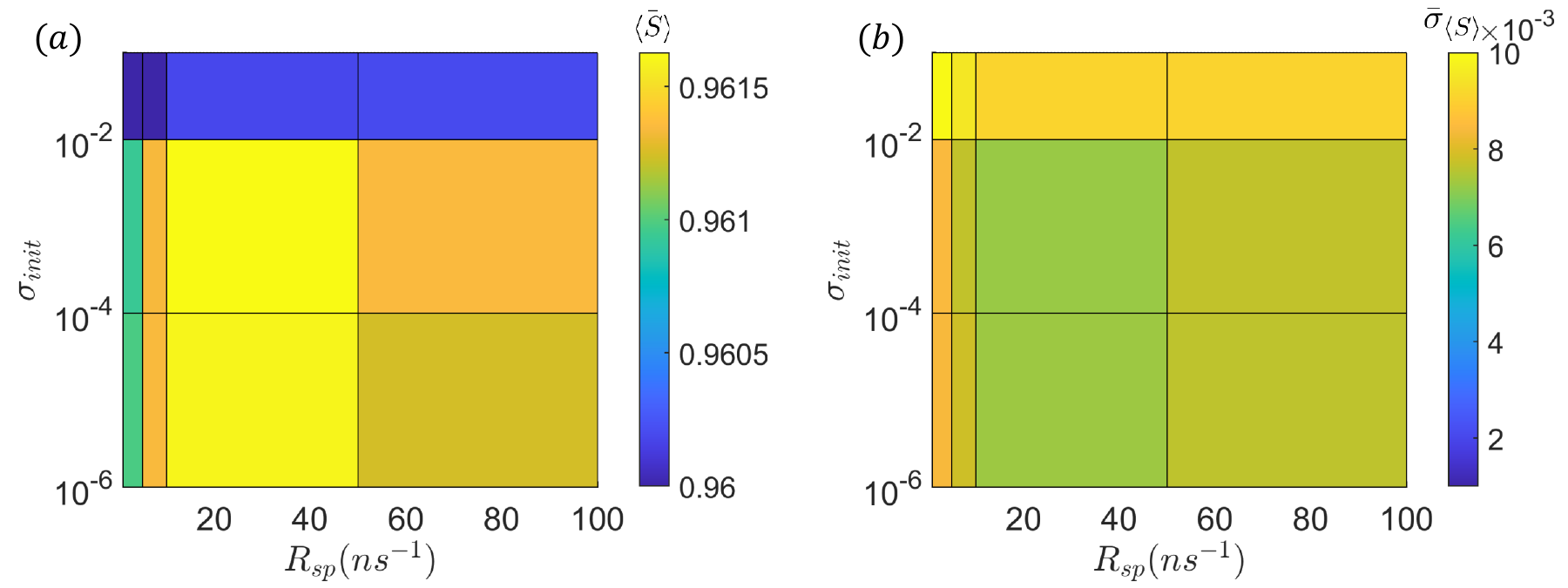}
\caption{Robust synchronization against dynamic and initial-condition noises. The optimal sparse laser network is from Fig.~\ref{fig:connect_scaling}(A) in the main text, obtained from the island-based genetic algorithm. (a,b) Color-coded plot of the average synchronization strength $\bar{\langle S\rangle}$ and the standard deviation of $\langle S\rangle$, respectively, in the 2D parameter plane of $(R_{\text{sp}},\sigma_{\text{init}})$, where four values are chosen for each parameter: $R_{\text{sp}} = \{5, 10, 50, 100\}\,\textnormal{ns}^{-1}$ and $\sigma_{\text{init}} = \{10^{-6}, 10^{-4}, 10^{-2}, 10^{-1}\}$. For each combination of $\{\sigma_{\text{init}}, R_{\text{sp}}\}$, $\langle S \rangle$ is calculated using 100 independent realizations, which are used to calculate the the quantities $\bar{\langle S \rangle}$ and $\bar{\sigma}_{\langle S \rangle}$.}
\label{fig:supp_noise_robust}
\end{figure}

In addition to the noises due to spontaneous emission and carrier-number fluctuations, the initial conditions in the time-delay buffer are also subject to stochastic fluctuations, giving rise to another source of Gaussian noise in the complex electrical field and carrier number:
\begin{align} \label{eq:ic_noise}
\Re[E^{s}_{p}(t)](1+\sigma_{\textnormal{init}}\mathcal{N}(0,1)), \,\Im[E^{s}_{p}(t)](1+\sigma_{\textnormal{init}}\mathcal{N}(0,1)),\,N_s(1+\sigma_{\textnormal{init}}\mathcal{N}(0,1)).
\end{align}
In our simulations, the Gaussian random variables in Eq.~[\ref{eq:ic_noise}] are independently generated for each laser and for each point in the initial time-delay interval. We integrate Eqs.~[\ref{eq:LK_noise_1}] and [\ref{eq:LK_noise_2}] using MATLAB’s \texttt{dde23} solver~\cite{shampine:2001} in the noise-free case and employ the Adams–Bashforth–Moulton stochastic integration method~\cite{toral:2014} for the case with dynamic noise. 

Our goal is to assess how robust the optimal sparse network in the main text is against the dynamic and initial-condition noises by calculating the average synchronization strength $\bar{\langle S\rangle}$ in the 2D parameter plane of ($R_{sp}, \sigma_{\textnormal{init}}$). The results are shown in Figs.~\ref{fig:supp_noise_robust}(a) and \ref{fig:supp_noise_robust}(b). Even under 10\% initial-condition noise ($\sigma_{\textnormal{init}}=0.1$) and relatively large spontaneous emission noise ($R_{sp}=100\,ns^{-1}$), stable and strong synchronization can be achieved, with fluctuations in the synchronization strength within about 0.2\%. The results in Figs.~\ref{fig:supp_noise_robust}(a) and \ref{fig:supp_noise_robust}(b) indicate that the optimal sparse networks found by the genetic algorithm can achieved robust synchronization against different types of noises in the diode lasers.

\subsection{Robustness against random variations in frequency detuning, coupling strength, and pump rate} \label{subsec:SI:Robust:FreqCoupPump}

\begin{figure} [ht!]
\centering
\includegraphics[width=0.8\linewidth]{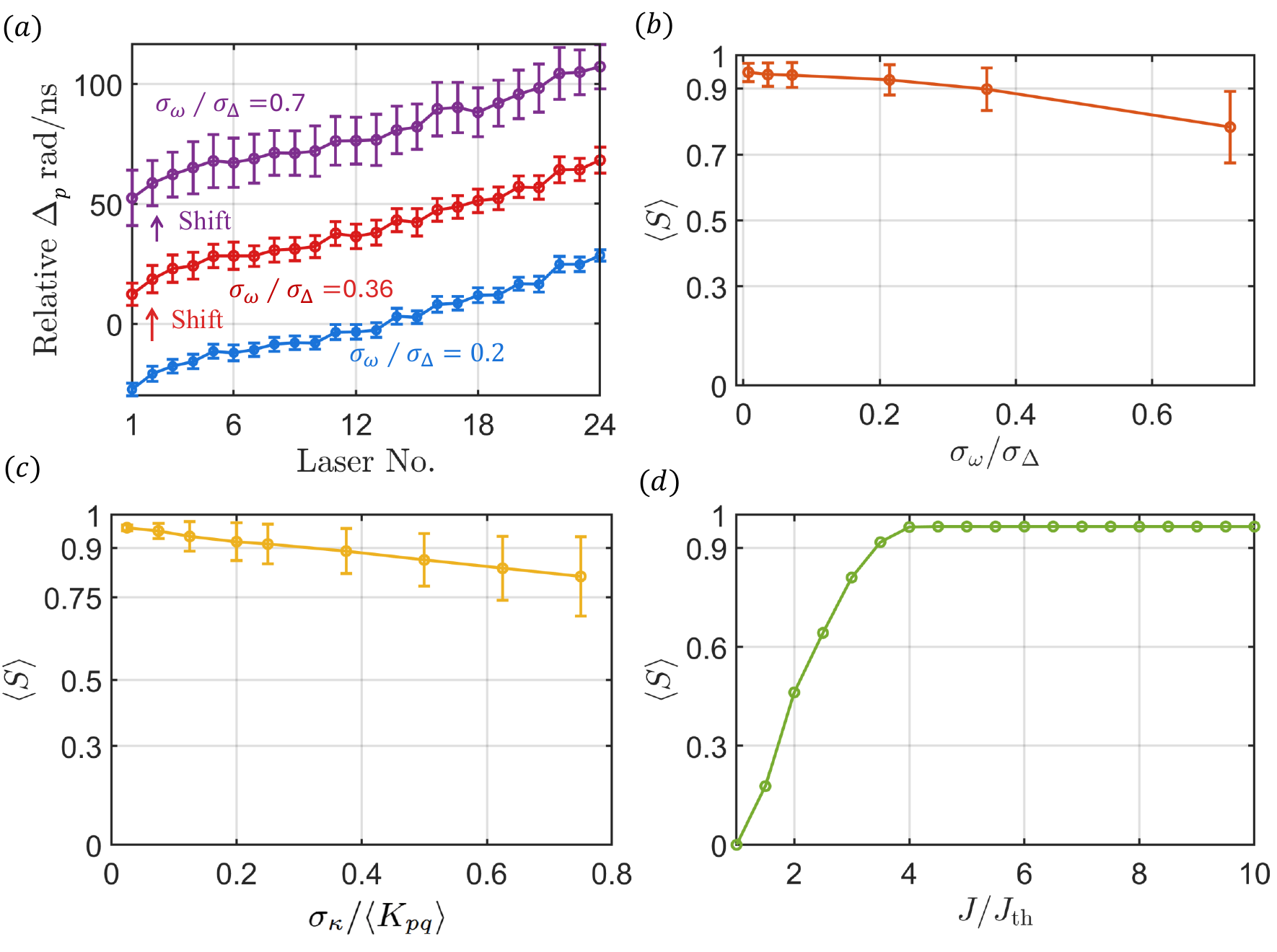}
	\caption{Robustness in synchronization of the optimal sparse network against perturbations in frequency disorder, nodal coupling strength, and pump rate with the initial frequency standard deviation of $\sigma_{\Delta}=14\,{\rm rad/ns}$. (a) Perturbation applied to the frequency detuning: $\Delta_p \rightarrow \Delta_p + \delta\omega_p$, where $\delta\omega_p = \sigma_{\omega}\, \mathcal{N}(0,1)$, with $\mathcal{N}(0,1)$ denoting a standard Gaussian random variable of zero mean and unit variance. Shown is the relative frequency detuning, shifted by $40$ rad/ns, for $\sigma_{\omega} = 3,\, 5$ and $10$ rad/ns, with standard error bars calculated from 100 random realizations. (b) Resulting synchronization strength as a function of $\sigma_{\omega}$ for $\sigma_{\omega} = 0.1, 0.5, 1, 3, 5, 10\,\textnormal{rad/ns}$, normalized by $\sigma_{\Delta}$. Error bars denote the standard deviation over 100 random realizations at each $\sigma_{\omega}$. (c) Synchronization strength $\langle S\rangle$ under perturbations to the nodal coupling strength: $K_{pq}\rightarrow K_{pq}+\delta K_{pq}$, where $\delta K_{pq} = \sigma_{\kappa}\, U(0,1)$, for $\sigma_{\kappa} = 0.01,0.03,0.05,0.08,0.1,0.15,0.2,0.25,0.3\,\textnormal{ns}^{-1}$. The mean coupling value is obtained by $\langle K_{pq}\rangle=\sum_{pq}K_{pq}/L^{2}\approx \chi^{*}$. (d) Threshold pump rate $J_{\rm th}=\gamma_n(N_0+\gamma/g)$. The pump is scanned over $J/J_{\rm th}\in[1,10]$, and the corresponding synchronization measure $\langle S\rangle$ is plotted as a function of $J/J_{\rm th}$. In the present study, the operating pump is $J_0=4J_{\rm th}\approx 3.67\times10^8\,\textnormal{ns}^{-1}$.}
\label{fig:supp_perb_detuning_coupling}
\end{figure}

The optimal sparse network in Fig.~\ref{fig:hub} in the main text was obtained for a fixed set of frequency detuning disorders. What is the effect of random fluctuations in the frequency detuning on synchronization? To answer this question, we perturb the frequency disorder $\Delta_p$ to $\Delta_p+\sigma_{\omega}\mathcal{N}(0,1)$, where $\mathcal{N}(0,1)$ denotes a standard Gaussian random variable of zero mean and unit variance, as illustrated in Fig.~\ref{fig:supp_perb_detuning_coupling}(a). Remarkably, even when the perturbation strength $\sigma_{\omega}$ reaches approximately 35\% of the original disorder strength $\sigma_{\Delta}$, the synchronization strength $\langle S\rangle$ decreases only modestly: from 0.96 to 0.90 (a mere 6.7\% reduction) as shown in Fig.~\ref{fig:supp_perb_detuning_coupling}(b). This result suggests that, insofar as the relative ordering of the frequency disorder magnitudes is preserved as shown in Fig.~\ref{fig:supp_perb_detuning_coupling}(a), strong frequency and phase synchronization associated with the optimal sparse network can be robustly maintained.

What is the effect of random perturbations in the nodal coupling strengths of the optimal sparse network on synchronization? To answer this question, we introduce perturbations to the coupling matrix in the form $K_{pq}+\sigma_{\kappa}U(0,1)$, where $U(0,1)$ denotes a uniform random variable, as shown in Fig.~\ref{fig:supp_perb_detuning_coupling}(c), and the perturbations make the coupling configuration no longer binary. We find that the synchronization strength decreases approximately linearly as the perturbation strength $\sigma_{\kappa}$ increases. The mean coupling is given by $\langle K_{pq}\rangle=\sum_{pq}K_{pq}/L^{2}\approx \chi^{*}$ in the optimized sparse coupling network. As shown in Fig.~\ref{fig:supp_perb_detuning_coupling}(c), a 25\% perturbation by $\sigma_{\kappa}/\langle K_{pq}\rangle$ leads to a comparable reduction in synchronization strength from 0.96 to 0.90 (a 6.7\% reduction).

The pump rate also influences the degree of synchronization. When the pump rate is too low, the electric field is insufficiently excited, hindering synchronization. Conversely, when the pump rate is excessively high, the electric field saturates, offering no additional benefit to synchronization. In our analysis, the pump rate is evaluated in normalized form, $J/J_{\rm th}\in[1,10]$. As the pump rate decreases from its original value [$J = 4J_{\rm th}$ in our study], synchronization tends to decline; however, a 13\% reduction still yields a high synchronization level with $\langle S\rangle>0.9$. In the opposite direction, increasing the pump rate beyond the original value leads to a saturation effect, with synchronization remaining nearly constant at higher pump levels.

Taken together, the results in Fig.~\ref{fig:supp_perb_detuning_coupling} demonstrate the synchronization robustness of the optimal sparse network against perturbations in the frequency disorder, the nodal coupling strength and pump rate.

\subsection{Optimization landscape for island genetic algorithm} \label{sec:SI:Robust:GA}

In the main text, the random frequency disorder is generated according to $\Delta = 14 \times \mathcal{N}(0,1)\,\textnormal{rad/ns}$ with the random seed \texttt{rng(1)} for a network of $L=24$ lasers. Each independent run of the island genetic algorithm produces a distinct optimal sparse matrix that defines the network. Do these different optimal networks still possess a similar synchronization capability/behavior? As the optimal solutions correspond to some valley (peak) or minima (maxima) on the optimization landscape, equivalently one can ask whether the optimal networks (matrices) lie within the same basin of 
attraction~\cite{langhojer:2005}? 

\begin{figure} [ht!]
\centering
\includegraphics[width=\linewidth]{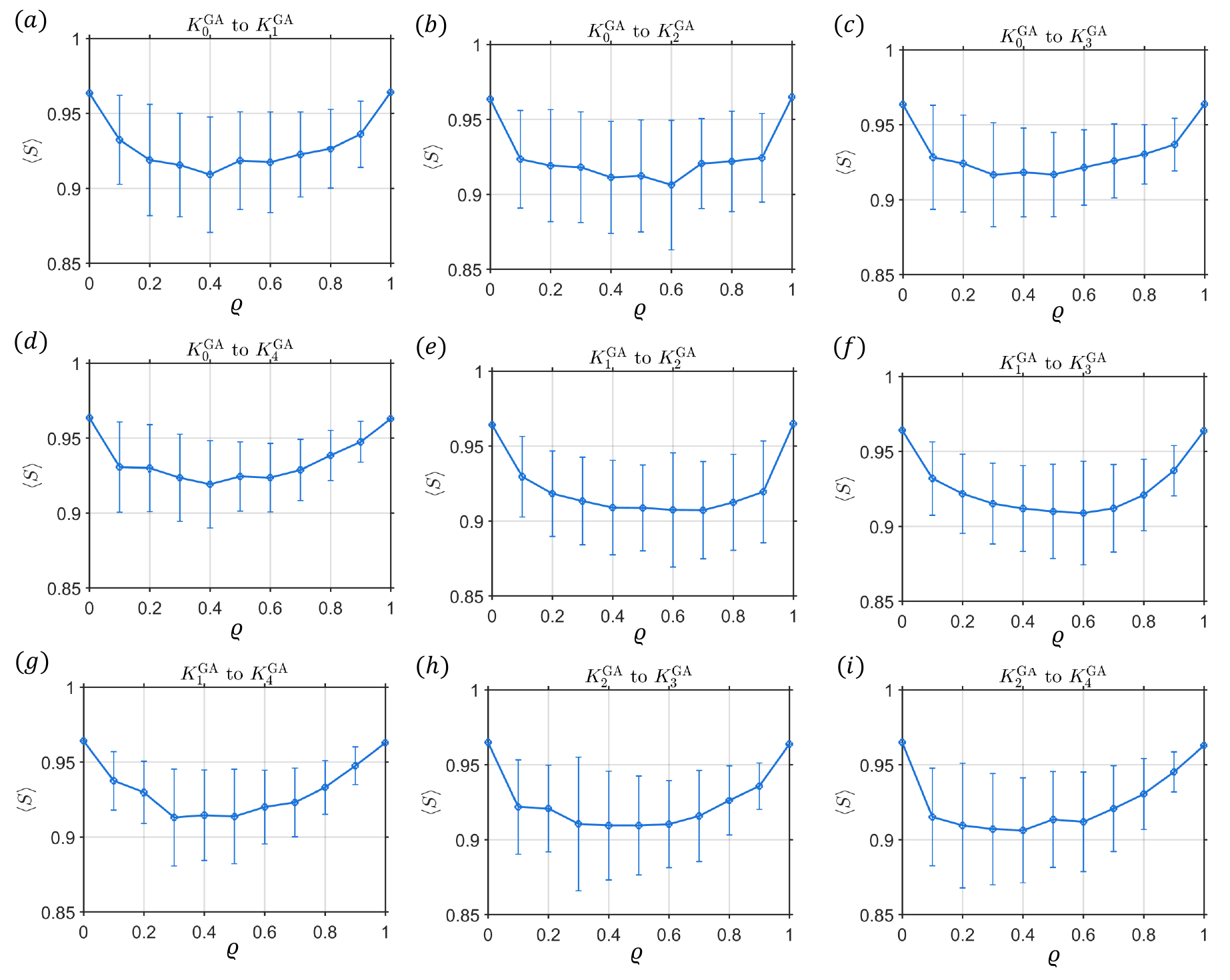}
\caption{Demonstration of the capability of island genetic algorithm in generating near optimal solutions of sparse network matrices. (a-i) Synchronization strength $\langle S\rangle$ for ``mixed'' optimal sparse networks. Five optimal networks are first generated by the genetic algorithm for $L=24$ and $\Delta = 14 \times \mathcal{N}(0,1)\,\textnormal{rad/ns}$ (from random seed \texttt{rng(1)}). The coupling matrices are interpolated (or ``mixed'') using the mixing parameter $\varrho\in [0,1]$ - see text for details. For each value of $\varrho$, 100 random realizations are used to calculate the average $\langle S\rangle$ value and its standard deviation. The fact that the $\langle S\rangle$ values in all cases remain high with relatively small variations implies that the genetic algorithm is capable of exploring diverse regions of the solution space to produce the optimal sparse network.}
\label{fig:supp_basin}
\end{figure}

To address this question, we ``mix'' the optimal networks on a pairwise base and introduce a parameter $\varrho\in [0,1]$ to characterize the mixing. In particular, for a given frequency disorder configuration, five independent executions of the island-based genetic algorithm with identical hyperparameters but different random seeds yield five optimal coupling matrices, denoted as $K^{GA}_{a}$ for $a=0,1,2,3,4$. Since each matrix is required to be binary, a direct interpolation is not applicable. Instead, for any pair of matrices $K^{GA}_a$ and $K^{GA}_b$, we define the difference matrix as $\Delta K^{GA}_{ab}=K^{GA}_a-K^{GA}_b$, whose zero elements correspond to the connections shared by the two matrices. To quantify the difference, we count the number of elements in the upper triangular part of $\Delta K^{GA}_{ab}$ with values equal to $-1\,ns^{-1}$ and $1\,ns^{-1}$, denoted as $N^{-1}_{ab}$ and $N^{1}_{ab}$, respectively. Since all five optimal coupling matrices have the same connectivity level, the total coupling cost is identical across them, meaning that the total numbers of edges of the five matrices are the same, implying $N^{-1}_{ab} = N^{1}_{ab}$. In fact, $N^{-1}_{ab}$ is the number of links absent in $K^{GA}_a$ but present in $K^{GA}_b$, i.e., $\{(p,q)|\Delta K^{GA}_{ab}(p,q)=-1\,ns^{-1}\}$, with $N^{1}_{ab}$ corresponding to the links present in $K^{GA}_a$ but absent in $K^{GA}_b$. In the mixing process, we retain all the common links (i.e., those with $\Delta K^{GA}_{ab}=0$) but modify the rest by randomly selecting a fraction $\varrho$ of the distinct links. Specifically, we randomly add $\varrho N^{-1}_{ab}$ links from the set in which $K^{GA}_{b}$ has connections but $K^{GA}_{a}$ does not, and remove $\varrho N^{1}_{ab}$ links from the set where $K^{GA}_a$ has connections but $K^{GA}_b$ does not. The resulting mixed matrix constructed from $K^{GA}_a$ and $K^{GA}_{b}$, denoted as $K^{GA}_{\textnormal{mix}}$, preserves the properties of being binary and symmetric with zero-diagonal constraints as described in the main text. By this construction, for $\varrho=0$, we recover $K^{GA}_{\textnormal{mix}}=K^{GA}_a$ and, for $\varrho=1$, we have $K^{GA}_{\textnormal{mix}}=K^{GA}_b$.

The optimization landscape revealed by this mixing process is illustrated in Figs.~\ref{fig:supp_basin}(a–i), where pairwise mixtures are performed among five independently optimal sparse coupling matrices. Each matrix differs from the others by approximately 30\% of the links, yet all leading to similarly high synchronization performance, indicating that the genetic algorithm has effectively identified some local optimum with a relatively large basin of attraction. Note that, if the level of consistency in synchronization decreased, the mixed network configurations would reside in distinct valleys of a rugged, non-convex optimization landscape. Thus, even if there are multiple, well-separated basins of attraction, the genetic algorithm is capable of exploring diverse regions of the solution space to produce the optimal sparse network, in spite of the lack of guarantee that the algorithm would find the global optimum.

While the global optimum is unknown, the island genetic algorithm consistently discovers distinct, high-quality local optima in the sense of near-maximal synchronization, highlighting the algorithm’s robustness and its ability to perform reliably in complex, non-convex combinatorial optimization problems. We find that, in the worst case of the mixing process, the synchronization measure remains about $\langle S\rangle=0.9$, corresponding to only a 5\% reduction from the best achievable value, further illustrating the robustness of the optimization landscape.

\begin{figure} [ht!]
\centering
\includegraphics[width=\linewidth]{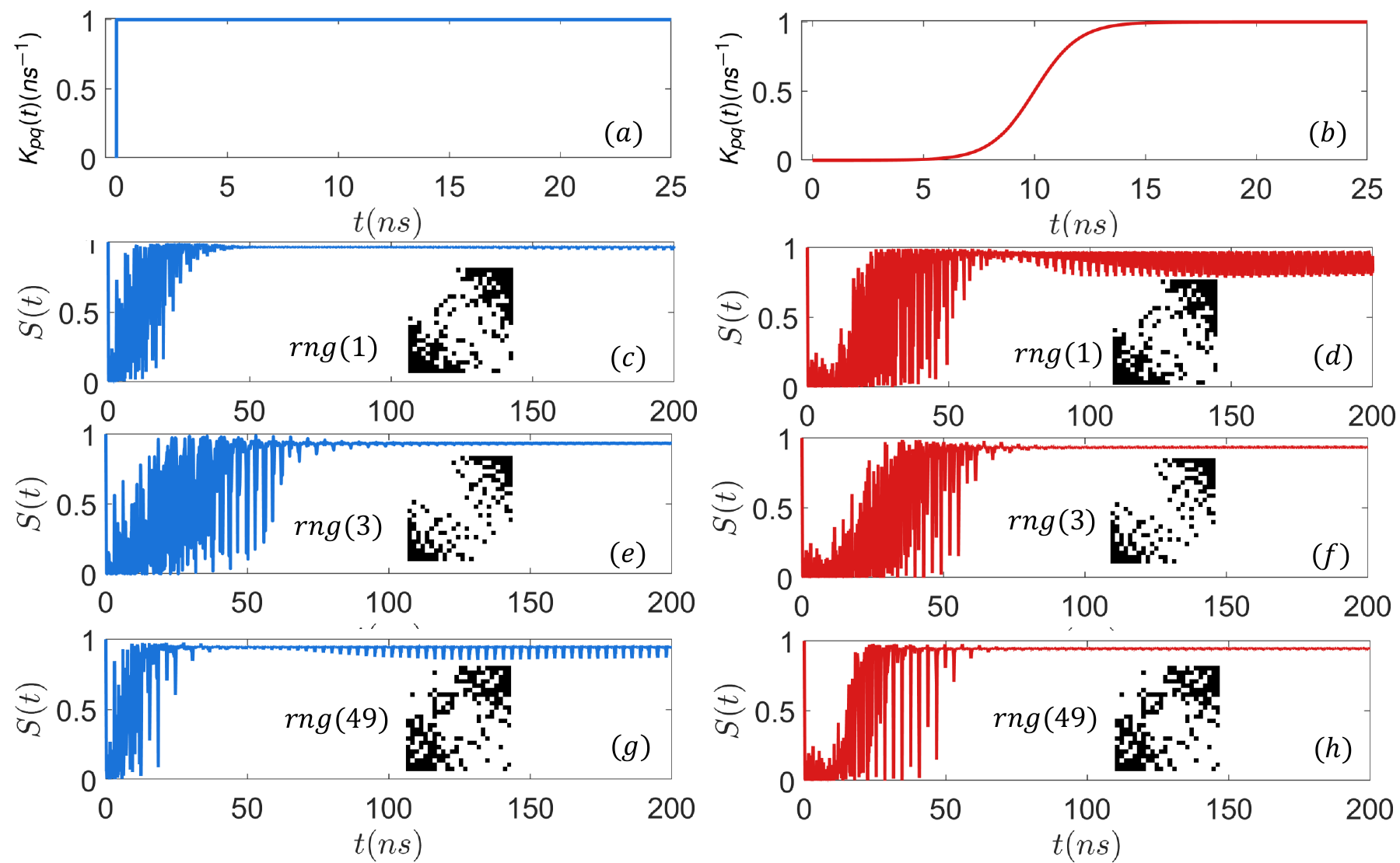}
\caption{Synchronization comparison between stepwise and smooth coupling ramps. Panels (a,b) illustrate two strategies for ramping the coupling strength, discontinuous (stepwise) and continuous via ($K_{pq}(t)=K_{pq}/[1+\exp[-[t-10]]]$), with the same target coupling configuration. Panels (c–h) report synchronization outcomes for three distinct frequency-disorder realizations (labeled by \texttt{rng}). Discontinuous ramps appear in (c,e,g), while continuous ramps appear in (d,f,h). In each pair (c,d), (e,f), and (g,h), the final coupling configuration after tens of nanoseconds is the same and fixed.}
\label{fig:supp_two_coupling_ways}
\end{figure}

\subsection{Synchronization under time-dependent coupling ramping} \label{subsec:SI:Robust:ramping}

In our simulations so far, a discontinuous coupling profile over time is assumed, where the strength of each coupling element is turned on abruptly to a nonzero constant at $t=0$, as shown in Fig.~\ref{fig:supp_two_coupling_ways}(a). Will a smooth, continuous coupling ramp, as shown in Fig.~\ref{fig:supp_two_coupling_ways}(b), affect synchronization? Figures~\ref{fig:supp_two_coupling_ways}(c–h) show, for $L=24$, the time evolution of synchronization under several optimized sparse coupling configurations with a hub structure. These results indicate that, for a given coupling topology, different ramping schemes can have no significant effect on the final synchronization, in spite of small differences in the dynamical behavior of the synchronization measure. Overall, the choice of the way by which coupling is turned on has little effect on the final synchronization. 

\section{Coupled network of linear oscillators} \label{sec:SI:LinearNetwork}

We provide a qualitative understanding of the optimal sparse network topology with a simple model of coupled linear oscillators. Briefly, neglecting gain saturation, $\alpha$ factor and time-delay of the coupling, we calculate the eigenstates (supermodes) of the coupled lasers with frequency disorder~\cite{nair:2019almost}. With increasing coupling, the eigenstates are no longer localized to individual lasers, but expand over multiple lasers. The eigenstate with the highest eigenvalue dominates lasing and its order parameter can be calculated. For selective sparse coupling, only the laser pairs with natural frequency differences above a threshold $\Delta_{\rm th}$ are coupled with a constant coefficient, while the others are not coupled at all. With increasing $\Delta_{\rm th}$, the network connectivity $\chi$ decreases. The eigenstates can be compared to those of randomly connected networks with the same value of connectivity $\chi$. With increasing $\chi$, the order parameter of the dominant supermode increases much more quickly for selective coupling than random coupling. We also find that the scaling of the critical connectivity follow the same trend. The qualitative agreement with the LK simulation results suggest that the linear coupling matrix determines the supermode structure, which in term dictates laser synchronization. However, this simple model cannot predict the destabilization of synchronization at even stronger coupling where the coupled lasers chaotically hop between multiple supermodes~\cite{davidchack2000chaotic,soriano2013complex}. 

Specifically, we examine whether sparse and selective coupling enables faster global phase and frequency synchronization, while requiring lower total coupling strength than conventional all‑to‑all or nearest‑neighbor configurations. To build design intuition and isolate the effect of coupling topology from intrinsic laser nonlinearities, we model the system as a network of linearly coupled oscillators that represent the laser array:
\begin{equation} \label{eq:linmodel}
    \frac{dE_p(t)}{dt} = (i \Delta_p - \gamma) E_p(t) + \sum^L_{q\neq p} K_{pq}E_q(t),
\end{equation}
where $E_p(t)$ denotes the slowly varying complex field amplitude of laser $p$, and $\Delta_p$ is its detuning relative to a rotating frame at $\omega_0$. Each cavity has a loss rate of $\gamma=500~{\rm ns^{-1}}$. The coupling topology is specified by the symmetric, sparse, binary matrix $K$, whose off‑diagonal elements represent the coupling strength between lasers $p$ and $q$ ($K_{pq} = K_{qp}$, either 0 or $\kappa=1~\rm{ns^{-1}}$); diagonal entries are set to zero to exclude self‑feedback. The overall network connectivity is quantified as
\begin{align}
\chi = \frac{1}{L(L-1)} \sum_{pq} K_{pq} / \kappa.
\end{align}
This linear system serves as a simplified representation of the Lang-Kobayashi model, neglecting gain dynamics, phase-amplitude coupling, and propagation delay. While it does not yield quantitative predictions, it provides clear intuition for designing coupling topologies.

\begin{figure} [ht!]
\centering
\includegraphics[width=1\linewidth]{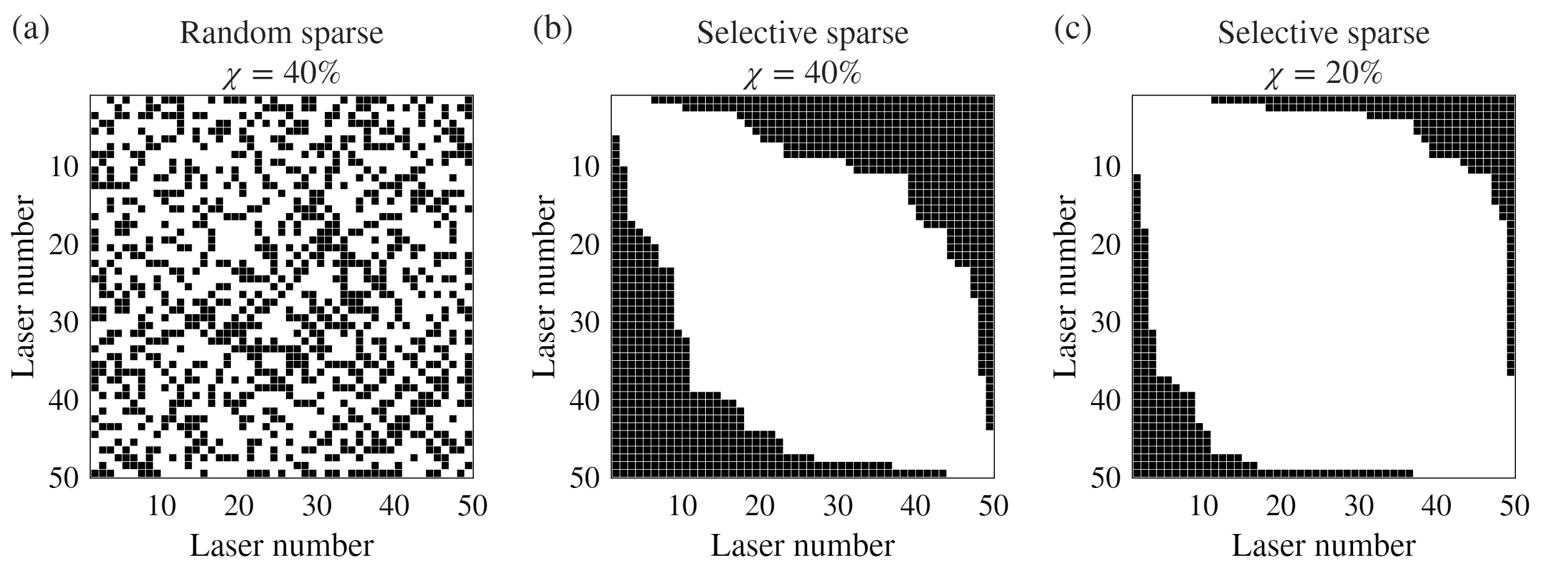}
\caption{Example of the different coupling matrix for an array of $L=50$ lasers. The laser are ordered from the most negatively detuned (laser 1) to the most positively detuned (laser 50). A black square indicate a link with coupling strength $\kappa = 1~\rm{ns^{-1}}$. (a) Random sparse matrix with a connectivity of $\chi = 40~\%$. The link are symmetric and random distributed over all laser. (b) and (c) are example of the Selective sparse matrix for 40 \% and 20 \% connectivity respectively. In this configuration the most detuned laser are connected first until the required connectivity is achieved.}
\label{fig:supp_linmodel_matrix}
\end{figure}

The formal solution of Eq.~[\ref{eq:linmodel}] can be expressed as $E^{l}_p(t) = v^{l}_p e^{\lambda^{l} t}$, where $\lambda^{l}$ is generally a complex eigenvalue, $p$ denotes the laser index ($p=1,2,\ldots,L$), and $l$ denotes the eigenvalue index. Substituting this ansatz into Eq.~[\ref{eq:linmodel}] gives
\begin{equation}
\lambda^{l} v^{l}_p=(i\Delta_p-\gamma)v^{l}_p+\sum_{q\neq p}^{L}K_{pq}v^{l}_q,
\end{equation}
which leads to the eigenvalue equation
\begin{equation}
\sum_{q\neq p}^{L}A_{pq}v^{l}_{q}=\lambda^{l} v^{l}_p,
\end{equation}
or equivalently, $\mathbf{A} \cdot \mathbf{v}^{l}=\lambda^{l} \mathbf{v}^{l}$. The matrix $\mathbf{A}$ takes the form
\begin{equation} \label{eq:linmodel-eigenproblem}
	\mathbf{A} = \begin{bmatrix}
(i\Delta_{1}-\gamma) & K_{12} & \cdots & K_{1L}\\
K_{21} & \ddots & ~ & \vdots\\
\vdots & ~ & \ddots & K_{(L-1)L}\\
K_{L1} & \cdots & K_{L(L-1)} & (i\Delta_{L}-\gamma)
\end{bmatrix},
\end{equation}
which encodes both the coupling matrix $K$ and the intrinsic dynamics of the individual oscillators, $(i\Delta_p - \gamma)$.

The eigenvalue spectrum $\lambda^{l}$ and the corresponding eigenvectors $\mathbf{v}^{l}$   of the matrix $\mathbf{A}$ in Eq.~[\ref{eq:linmodel-eigenproblem}] determine the behavior of the network’s ``supermode'' solutions. Because all eigenvalues possess negative real parts, the eigenvalue with the largest real component identifies the dominant or slowest-decaying supermode. Consequently, the analysis focuses on the corresponding eigenvector $\mathbf{v}^{l_{\rm max}}$, associated with the eigenvalue having the largest ${\rm Re}[\lambda^{l_{\rm max}}]$. For compactness, the field of laser $p$ is written as $E_p(t)=v_pe^{\lambda t}$. The imaginary part of $\lambda$ specifies the oscillation frequency of the collective mode.

Each field $E_p(t)$ carries both the intrinsic phase of laser $p$ and a global phase contribution from $e^{\lambda t}$.  The synchronization among lasers is quantified using the following order parameter
\begin{equation}
S=\frac{\left| \sum_{p} v_p \right|^2}{L \sum_{p}\left| v_p \right|^2},
\end{equation}

Since $e^{\lambda t}$ is a common time-dependent factor only the complex amplitude $v_p$ of the field $E_p(t)$ matters. To investigate how the dominant supermode attains a high degree of phase synchronization—specifically, how individual lasers within the network synchronize under varying coupling topologies—we simulate two types of sparse coupling configurations for laser arrays with sizes ranging from $L=50$ to $L=200$. The first configuration, referred to as the random sparse topology, is generated by randomly removing links from the full coupling matrix until the desired connectivity is achieved. The second configuration, termed the selective sparse topology, is obtained by progressively removing links between lasers with the smallest detuning values (see Fig.~\ref{fig:supp_linmodel_matrix} for representative matrices). For each topology, the connectivity $\chi$ is varied from  0\% to 100\%, and simulations are performed over 100 realizations of laser frequency detuning drawn from a normal distribution with a standard deviation of $\sigma = 14 ~{\rm rad / ns}$. In the random sparse case, an additional ensemble average is conducted over 10 distinct random realizations of the coupling matrix.

\begin{figure} [ht]
\centering
\includegraphics[width=1\linewidth]{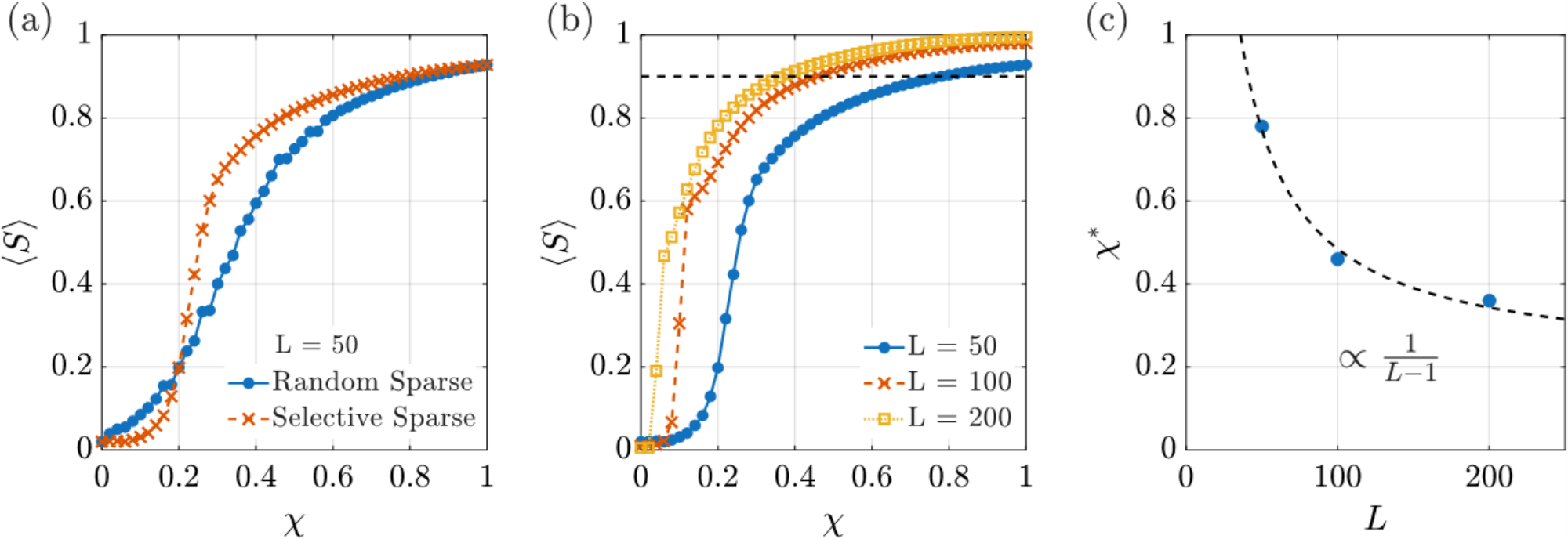}
\caption{Simulation results of the linear coupled network. (a) Average  order parameter $\langle S \rangle$ of the dominant super-mode $\mathbf{v}^{l_{\rm max}}$ as a function of connectivity $\chi$ for $L = 50$ lasers array. The blue dot are for a random sparse matrix and the red cross for a selective one. (b) Comparison of the average  order parameter $\langle S \rangle$ as a function of connectivity for different numbers of lasers in a selective sparse configuration. Blue dot is for $L = 50$, red cross $L = 100$ and yellow square for $L = 200$. The dashed black line is set to  $\langle S \rangle = 0.9$ and used to define the critical connectivity $\chi^*$. All results are the average of 100 normally distributed frequency realizations with a standard deviation of $\sigma = 14 ~\rm{rad/ns}$. The Random sparse configuration is also averaged over 10 different random configurations. (c) Scaling of the critical connectivity $\chi^*$ as a function of the number of laser $L$ in the array for a selective sparse configuration.}
\label{fig:supp_linmodel}
\end{figure}

Figure~\ref{fig:supp_linmodel}(a) compares the averaged synchronization order parameter $\langle S \rangle$ of the dominant supermode for both random and selective sparse configurations.  The selective sparse topology grows to a synchronized state faster than the random sparse network. This improved efficiency arises because, in the selective sparse case, the available coupling strength is directed primarily toward lasers with larger frequency detuning. Synchronization theory predicts that for nonidentical oscillators, the coupling must exceed the system’s intrinsic frequency spread to achieve phase locking~\cite{chopra:2009,jadbabaie:2004,dorfler:2011}. Concentrating coupling power on the most detuned lasers, as shown in Figs.~\ref{fig:supp_linmodel_matrix}(b–c), allows these elements to synchronize more rapidly and act as hubs that mediate global phase alignment across the entire array.

Figures~\ref{fig:supp_linmodel}(b) and \ref{fig:supp_linmodel}(c) further illustrate the scalability of the selective sparse topology. We define a critical connectivity $\chi^*$ as the point where $\langle S \rangle = 0.9$, indicating strong synchronization of the dominant eigenmode. Repeating the analysis for arrays with $L = 50,~100,200$ lasers reveals that $\chi^*$ decreases systematically with increasing $L$, following $\chi^* \propto 1/(L-1)$, consistent with the expected scaling behavior. Thus, larger arrays require fewer overall links to achieve global synchronization. However, if connectivity is reduced excessively in the selective sparse configuration, some lasers may become disconnected from the network. Therefore, an optimal balance between selective and random coupling should be employed for large-scale arrays.

\section{Effective thermodynamic potential for the laser network}
\label{sec:SI:Thermo}

The theoretical objective is to identify synchronized steady-state solutions, characterized by the same frequency and phase at the same constant amplitudes among lasers, in the presence of frequency disorders and arbitrary coupling topologies. First, we expand the LK equations around the steady-state solutions of each laser within the coupled network to derive a generalized time-delayed Kuramoto mode~\cite{yeung:1999}. By applying a slow-phase-evolution approximation~\cite{lenstra:1991} and the synchronized steady-state condition as $\Omega_p(t)=\Omega_q(t)$, we obtain the gradient flow
\begin{equation}\label{eq:gradient_flow}
    \dot{\eta}_p(t)=-\frac{1}{\tau}\frac{d}{d\eta_p(t)}U(\eta_p(t)),
\end{equation}
governed by the thermodynamic potential for each laser~\cite{NHBWB:2021}:
\begin{equation} U(\eta_p(t)) = \eta^{2}_p(t) - 2\mathcal{F}_p\eta_p(t) - 2\mathcal{K}_p\cos[\eta_p(t)+\phi],
\label{eq:thermodynamic_potential}
\end{equation}
where $\eta_p(t) =\Omega_p(t) - \Omega_p(t-\tau)$ represents the delay-induced phase difference, and $\mathcal{F}_p = \tau\Delta_p$ denotes the effective detuning, while $\mathcal{K}_p \equiv \tau\sqrt{1+\alpha^2}k_p^{\text{in}}$ is the effective coupling strength, with $k_p^{\text{in}} \equiv \sum_{q=1}^{L} K_{pq}$ representing the intrinsic coupling from other lasers to the $p$-th laser within the coupling matrix. The frustration phase term~\cite{sakaguchi1986soluble,daido1992quasientrainment}, $\phi = \arctan\alpha + \omega_0\tau$, depends on the linewidth enhancement factor $\alpha$ and $\omega_0\tau$ is set as $2n\pi$ with integer $n$. 

The time delay $\tau$ introduced in Eq.~[\ref{eq:gradient_flow}] is due to the time constant setting. We can also renormalize the final solution $\eta_p(t)$ by multiplying both sides by the time constant $\tau$ to make the time scale unitless in Eq.~[\ref{eq:gradient_flow}]. By doing so, both the effective detuning and effective coupling become independent of the time delay. 

\begin{figure} [ht!]
\centering
\includegraphics[width=0.8\linewidth]{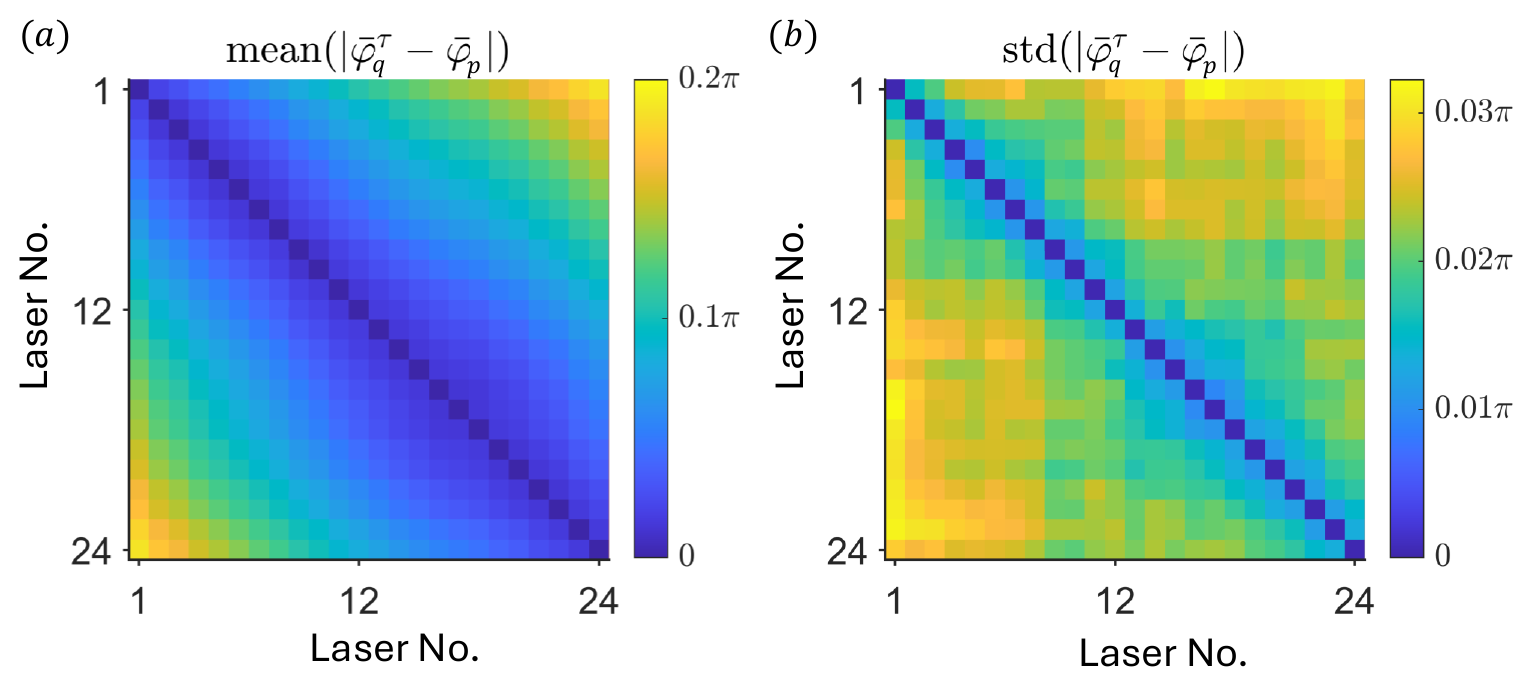}
\caption{The time-delayed phase differences in optimal sparse coupling matrices for nine different frequency-disorder realizations from Fig.~\ref{fig:supp_coupling}. (a-b) The mean and standard deviation of the time-delayed phase differences. Here, $\bar{\varphi}^{\tau}_q\equiv\bar{\varphi}_q(t-\tau)$ and $\bar{\varphi}_p\equiv\bar{\varphi}_p(t)$.}
\label{fig:supp_phase}
\end{figure}

The local minima of the potential $U(\eta_p(t))$ can be determined through the first and second derivatives with respect to $\eta_{p}(t)$:
\begin{align}
    \frac{dU(\eta_p(t))}{d\eta_p(t)}&=\eta_p(t)-\mathcal{F}_p+\mathcal{K}_p\sin[\eta_p(t)+\phi]=0,\label{eq: ECM_criteria_1}\\
    \frac{d^2 U(\eta_p(t))}{d\eta^2_p(t)}&=1+\mathcal{K}_p\cos[\eta_p(t)+\phi]>0.\label{eq: ECM_criteria_2}
\end{align}
The variational principle dictates that the system’s dynamical steady state described in Eq.~[\ref{eq:gradient_flow}] converges to a local minimum of the potential. Thus, the local minima correspond to the steady-state solutions for each individual laser under the coupled network. However, as the effective coupling strength increases, the number of possible steady‐state solutions also grows. In particular, the nonzero amplitude-phase coupling factor ($\alpha$), which acts as an intrinsic frustration parameter~\cite{sakaguchi1986soluble,daido1992quasientrainment}, destabilizes these steady states~\cite{winful1988stability,winful:1990,ye2026disorder}, leading to the emergence of multiple basins of attraction. The steady-state solutions from Eqs.~[\ref{eq: ECM_criteria_1}] and [\ref{eq: ECM_criteria_2}] represent the time‐delayed phase difference
\begin{equation}\label{eq: final_frequency_1}
    \eta_{p}(t) = \Omega^{fit}_p\tau + \varphi_p(t)-\varphi_p(t-\tau).
\end{equation}
In the case of target complete synchronization, the solutions of $\eta_p$ become independent of the laser index $p$ and reduce to
\begin{equation}\label{eq: final_frequency_2}
    \eta^{*}\approx \Omega^{fit}\tau+\textnormal{Const.}=2\pi f_{\textnormal{final}}\tau+\textnormal{Const.}
\end{equation}
with
\begin{equation}\label{eq:phase_approx}
    \varphi_p(t)-\varphi_p(t-\tau)\approx \textnormal{Const.}+2n\pi
\end{equation}
with integer $n$, supported by the numerical results in the Fig.~\ref{fig:supp_phase}. Therefore, the synchronized steady-state solutions of $\eta^{*}$ represent the final frequency values. 

The final optimized numerical results in Fig.~\ref{fig:supp_phase}(a,b) demonstrate the average and fluctuation of the time-delay phase difference as $|\bar{\varphi}_q(t-\tau)-\bar{\varphi}_p(t)|$ over the target time window, respectively.  Fig.~\ref{fig:supp_phase}(b) serve as the numerical support for Eq.~[\ref{eq:phase_approx}], with the synchronized steady-state condition to transform $|\bar{\varphi}_q(t-\tau)-\bar{\varphi}_p(t)|$ to $|\bar{\varphi}_p(t-\tau)-\bar{\varphi}_p(t)|$. Moreover, the Fig.~\ref{fig:supp_phase}(a) indicates that only the frequency outliers have a relatively large phase error relative to $2\pi$. This is the primary reason the final order parameter cannot reach perfect unity~\cite{mork1992chaos}, saturating instead at approximately $0.97$.

We note that the coupling coefficients $K_{pq}$ considered here are real-valued, with the common propagation phase included in the factor $e^{-i\omega_0\tau}$. In general, the coupling coefficients are complex-valued, $K_{pq}=|K_{pq}|e^{i\theta_{pq}}$. The coupling phase $\theta_{pq}$ appears in the reduced phase dynamics as an additional offset in the sinusoidal coupling function in Eq.~\ref{eq:thermodynamic_potential}, combining with the phase shifts associated with the time delay and the amplitude-phase coupling in $\phi$. In addition to coupling amplitudes, coupling phases constitute additional degrees of freedom that could be co-optimized with the network topology to further mitigate dynamical frustration.

\section{Physical understanding of the optimal sparse network structure}\label{sec:SI:theory}

The previous section presented the effective thermodynamic potential for the steady-state solutions of each laser coupled within the network. When one of these individual steady-state solutions completely overlaps, the system is in a synchronized steady state. This signifies that every laser converges to the same frequency and phase at a uniform constant amplitude.

In the uncoupled limit ($\mathcal{K}_p = 0$), the potential is purely quadratic with a unique global minimum, ensuring that any initial state is attracted to a single stable steady state. However, frequency disorder introduces deviations proportional to $\mathcal{F}_p$, shifting the steady-state solutions of individual lasers and preventing synchronization. The coupling term in Eq.~[\ref{eq:thermodynamic_potential}] counteracts these shifts by modulating the quadratic potential with oscillations whose oscillation frequency is high relative to the scale of the frequency disorder ($2\pi\ll\mathcal{F}_p$). The oscillation amplitude, governed by $\mathcal{K}_p$, determines both the values and the multiplicity of steady-state solutions. Unlike all-to-all coupling, which provides uniform coupling across the array, sparse networks introduce heterogeneity in $\mathcal{K}_p$. This variation allows the sparse architecture to prune extraneous solutions, thereby stabilizing the dynamics and aligning the steady-state values to achieve superior frequency and phase consistency, as shown in Fig.~\ref{fig:connect_scaling}(A) of the main text.

\begin{figure*} [ht!]
\centering
\includegraphics[width=0.8\linewidth]{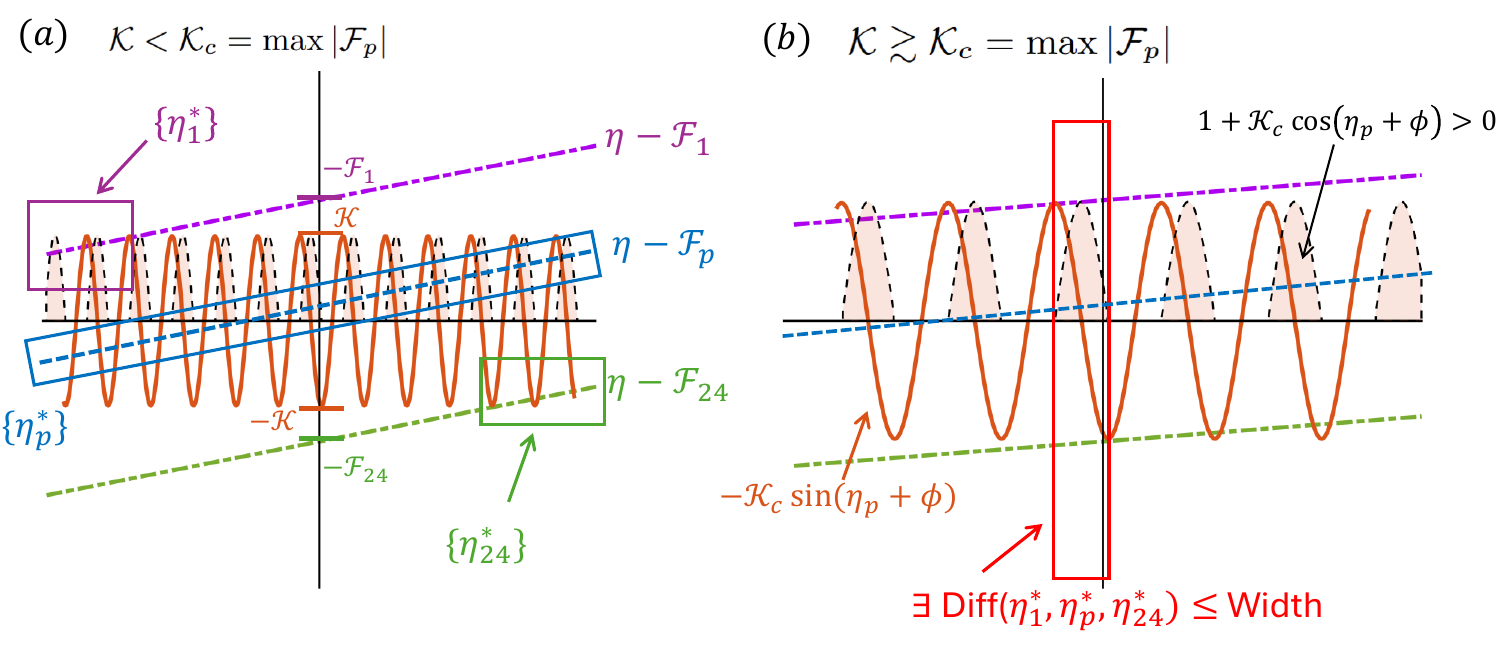}
\caption{Schematic of the global critical effective coupling. (a) Global effective coupling $\mathcal{K}$ below the critical value $\mathcal{K}_c$. (b) Global effective coupling $\mathcal{K}$ around the critical value $\mathcal{K}_c$. The intersections between the line $\eta_p - \mathcal{F}_p$ and the curve $-\mathcal{K}\sin[\eta_p + \phi]$ correspond to the solutions where $dU/d\eta = 0$. The shaded area represents the regions where $d^2U/d\eta^2 > 0$. The rectangular regions in (a) indicate the steady-state solutions for each laser with the corresponding effective detuning $\mathcal{F}_p$.The red rectangular region in (b) highlights the first nearly synchronized state that emerges at the critical effective coupling. Because the synchronized solutions for $\eta$ are proportional to the final frequency, the tolerance error of the first nearly synchronized states implies that the maximum frequency difference between lasers is approximately the width of the region satisfying $d^2U/d\eta^2 > 0$.}
\label{fig:supp_theory_critical_coupling}
\end{figure*}

Furthermore, we identify a global critical effective coupling, $\mathcal{K}_{c} = \max |\mathcal{F}_p|$, as shown in Figs.~\ref{fig:supp_theory_critical_coupling}(a,b). This is defined as the minimum global effective coupling required for the steady-state solutions of each laser to first nearly overlap with the tolerance error, as shown in Fig.~\ref{fig:supp_theory_critical_coupling}(b). It suggests that the global critical intrinsic coupling strength is given by
\begin{equation}
k_c^{\text{in}} = \max |\Delta_p| / \sqrt{1+\alpha^2}.
\end{equation}
Under the global critical effective coupling condition, the total coupling strength is
\begin{equation}
\sum_p k_c^{\text{in}} = \chi_c L(L-1)\kappa^{f} = L \max |\Delta_p| / \sqrt{1+\alpha^2},
\end{equation}
which defines the critical connectivity
\begin{equation}
\chi_c = \max |\Delta_p| / [\kappa^{f}\sqrt{1+\alpha^2}(L-1)] \propto 1/(L-1).
\end{equation} The optimal connectivity $\chi^*$ typically lies at the edge of chaos and is slightly larger than $\chi_c$. We thus have $\chi^* \gtrsim \chi_c \propto 1/(L-1)$, as shown in Fig.~\ref{fig:connect_scaling}(C) of the main text. 

\begin{figure} [ht!]
\centering
\includegraphics[width=0.8\linewidth]{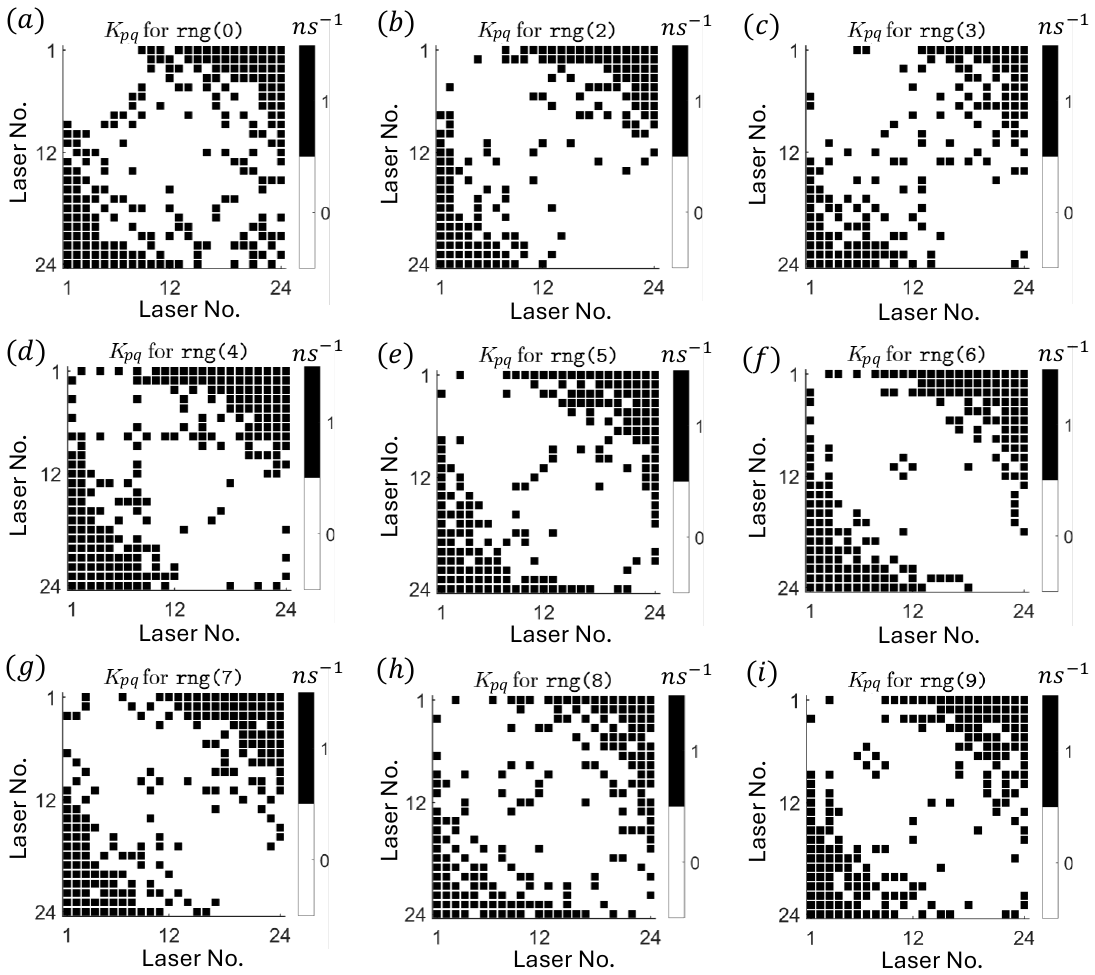}
\caption{Optimal sparse coupling matrices for nine different frequency-disorder realizations. (a-i) Binary color-encoded optimized coupling matrices corresponding to the frequency disorder $\Delta=14\times \mathcal{N}(0,1)\,\textnormal{rad/ns}$, generated using random seeds \texttt{rng(0)} to \texttt{rng(9)}, excluding \texttt{rng(1)}.}
\label{fig:supp_coupling}
\end{figure}

At the individual laser level within the sparse network, there exists a local critical effective coupling $\mathcal{K}_c^p = |\mathcal{F}_p|$ that guarantees the existence of a single steady-state solution near the initial mean frequency $\eta^*=0$, requiring
\begin{equation}
k_{c,p}^{\text{in}} = |\Delta_p| / \sqrt{1+\alpha^2}.
\end{equation}
The necessity for each laser to satisfy this individual threshold leads to the emergence of a hub structure within the optimal network, where lasers with larger local detuning require a greater number of connections to maintain synchronization, as shown in Fig.~\ref{fig:hub}(D) of the main text, and Fig.~\ref{fig:supp_coupling}. Figures \ref{fig:supp_coupling}(a–i) display the optimized sparse binary coupling matrices corresponding to different realizations of the frequency disorder. The row sum of each matrix, representing the number of connections of laser $p$, further supports the trends reported in the main text by illustrating how connectivity adapts to local detuning variations.

\begin{figure} [ht!]
\centering
\includegraphics[width=0.4\linewidth]{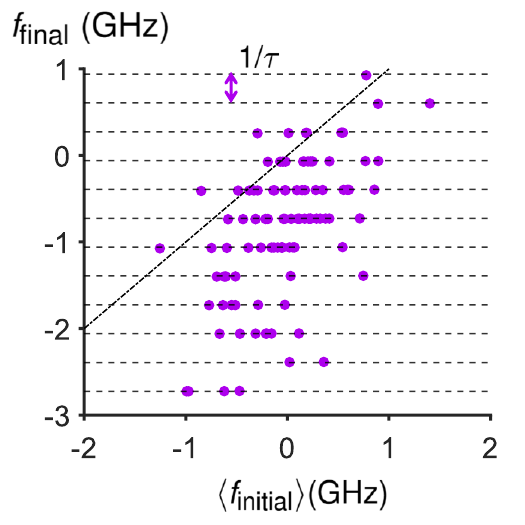}
\caption{Scatter plot of the final frequencies $f_{\rm final}$ of the synchronized lasers versus their initial mean $\langle f_{\rm initial}\rangle$ for 100 frequency-disorder realizations, corresponding to 100 data points, respectively. }
\label{fig:final_freq}
\end{figure}

The network optimization process is repeated with 100 statistically independent realizations of the frequency disorder, yielding $\langle S\rangle_e \approx 0.97 \pm 0.01$. As shown in Fig.~\ref{fig:final_freq}, the final frequencies of 100 synchronized networks, each optimized for its corresponding frequency disorder, are the discrete frequencies of external cavity resonances spaced by $1/\tau$ due to the external cavity feedback effect. The overall negative shift of the synchronized frequency relative to the mean of the initial frequencies is a consequence of the positive amplitude-phase coupling ($\alpha$ factor). Physically, a positive $\alpha$ implies that an increase in the laser intensity (driven by feedback coupling and coherent synchronization) leads to a reduction in carrier density and a corresponding increase in the refractive index of the active medium~\cite{henry2003theory}. This effectively increases the optical path length within the laser cavity, ultimately resulting in a frequency redshift~\cite{lang:1980}.

\FloatBarrier

\end{document}